\pdfoutput=1

\documentclass[11pt,twoside,a4paper,cmspaper,final,collab]{cms-tdr}
\def\svnVersion{55d0b26-D}\def\svnDate{2019/05/24}\def\cmsCernNoTag{CERN-EP-2018-221}\def\cmsCernDate{\today}\def\cmsMessage{"Published in the Journal of High Energy Physics as \href{http://dx.doi.org/10.1007/JHEP06(2019)093}{\doi{10.1007/JHEP06(2019)093}.}"}

\begin{document}\cmsNoteHeader{HIG-18-007}

\hyphenation{had-ron-i-za-tion}
\hyphenation{cal-or-i-me-ter}
\hyphenation{de-vices}
\ifthenelse{\boolean{cms@external}}{\providecommand{\cmsRight}{bottom\xspace}}{\providecommand{\cmsRight}{right\xspace}}
\providecommand{\CL}{CL\xspace}
\providecommand{\cmsTable}[1]{\resizebox{\textwidth}{!}{#1}}
\newlength\cmsTabSkip\setlength{\cmsTabSkip}{1ex}
\newcommand{\mvis}{\ensuremath{m_\text{vis}}\xspace}
\newcommand{\mtt}{\ensuremath{m_{\PGt\PGt}}\xspace}
\newcommand{\hww}{\ensuremath{\PH\to\PW\PW}\xspace}
\newcommand{\hzz}{\ensuremath{\PH\to\PZ\PZ}\xspace}
\newcommand{\htt}{\ensuremath{\PH\to\PGt\PGt}\xspace}
\newcommand{\VH}{\ensuremath{\text{V}\PH}\xspace}
\newcommand{\LT}{\ensuremath{L_{\text{T}}}\xspace}
\newcommand{\LTH}{\ensuremath{L_{\text{T}}^{\text{Higgs}}}\xspace}
\newcommand{\vST}{\ensuremath{\vec{S_{\text{T}}}}\xspace}
\newcommand{\mH}{\ensuremath{m_{\PH}}\xspace}
\newcommand{\zJets}{\ensuremath{\PZ+\text{jets}}\xspace}

\cmsNoteHeader{HIG-18-007}
\title{Search for the associated production of the Higgs boson and a vector boson in proton-proton collisions at $\sqrt{s}=13\TeV$ via Higgs boson decays to $\Pgt$ leptons}

\date{\today}

\abstract{
A search for the standard model Higgs boson produced in association with a $\PW$ or a $\PZ$ boson and
decaying to a pair of $\Pgt$ leptons
is performed. A data sample of proton-proton collisions collected at
$\sqrt{s} = 13\TeV$ by the CMS experiment at the CERN LHC is used, corresponding to an integrated luminosity of 35.9\fbinv.
The signal strength is measured relative to the expectation for the standard model Higgs boson, yielding
$\mu = 2.5 ^{+1.4} _{-1.3}$.
These results are combined
with earlier CMS measurements targeting Higgs boson decays to a pair of $\PGt$ leptons, performed with the same data set in
the gluon fusion and vector boson fusion
production modes.
The combined signal strength is $\mu = 1.24 ^{+0.29} _{-0.27}$
($1.00 ^{+0.24} _{-0.23}$ expected), and the
observed significance is 5.5 standard deviations (4.8 expected) for a Higgs boson mass of 125\GeV.
}

\hypersetup{
pdfauthor={CMS Collaboration},
pdftitle={Search for the associated production of the Higgs boson and a vector boson in proton-proton collisions at sqrt s = 13 TeV via Higgs boson decays to tau leptons},
pdfsubject={CMS},
pdfkeywords={CMS, physics, Higgs, Higgs associated production}}

\maketitle

\section{Introduction\label{sec:intro}}

In the standard model (SM), the fermions receive mass via their Yukawa couplings to the Higgs
boson~\cite{Englert:1964et,Higgs:1964ia,Higgs:1964pj,Guralnik:1964eu,Higgs:1966ev,Kibble:1967sv,ATLASobservation125,CMSobservation125,CMSobservation125Long},
and measurements
of the Higgs boson branching fractions to fermions directly probe these couplings.
The Higgs boson decay to a $\PGt$ lepton pair is particularly interesting because it
has the largest branching fraction among the direct leptonic
Higgs boson decays ($\mathcal{B}(\PH \to \PGt^+\PGt^-)\simeq 6.3\%$).
Many searches for the $\PH \to \PGt^+\PGt^-$ process have been performed by earlier
experiments~\cite{Barate:2000ts,Abdallah:2003ip,Achard:2001pj,Abbiendi:2000ac,Aaltonen:2012jh,Abazov:2012zj}.
The ATLAS and CMS Collaborations each previously reported evidence for this particular
Higgs boson decay process using data collected at center-of-mass energies of 7 and
8\TeV~\cite{SMTauTauRun1ATLAS,Aad:2015zrx,Chatrchyan:2014nva}.
The $\PH \to \PGt^+\PGt^-$ process was measured targeting the gluon fusion and vector boson fusion production
modes using data collected by the CMS Collaboration at a center-of-mass energy of 13\TeV~\cite{Sirunyan:2017khh}
resulting in a cross section times branching fraction of $1.09^{+0.27}_{-0.26}$ relative
to the SM expectation.

This paper reports on a search for the SM Higgs boson
produced in association with a $\PW$ or a $\PZ$ boson.
The Higgs boson is sought in its decay to a pair of $\Pgt$ leptons.
The search is based on a data set of proton-proton ($\Pp\Pp$)
collisions, collected in 2016 by the CMS experiment at a center-of-mass energy of $\sqrt{s}=13$\TeV,
corresponding to an integrated luminosity of 35.9\fbinv.
The results are combined
with prior results from the CMS $\PH \to \PGt^+\PGt^-$ analysis performed with the same data set and focusing on
the gluon fusion and vector boson fusion production modes~\cite{Sirunyan:2017khh}.
This combination provides dedicated signal regions covering the four leading Higgs
boson production mechanisms: gluon fusion, vector boson fusion, $\PW$ associated production, and $\PZ$ associated production.

For the $\PZ\PH$ associated production channel,
$\PZ\to \ell^+\ell^-$ ($\ell=\Pe,\Pgm$) decays are considered, combined with four possible $\PGt\PGt$
final states from the Higgs boson decay: $\Pe\tauh$, $\Pgm\tauh$,
$\Pe\Pgm$, and $\tauh\tauh$, where $\tauh$ denotes $\PGt$ leptons decaying hadronically. For the $\PW\PH$ channel, four final states are considered, with
the $\PW$ boson decaying leptonically to a neutrino and an electron or a muon (listed first in the 
following notation), and the Higgs boson decaying to at least one $\tauh$ (listed second):
$\Pgm+\Pgm\tauh$, $\Pe+\Pgm\tauh$/$\Pgm+\Pe\tauh$, $\Pe+\tauh\tauh$, and $\Pgm+\tauh\tauh$. The final state with an
electron, a muon, and a $\tauh$ candidate is written as $\Pe+\Pgm\tauh$/$\Pgm+\Pe\tauh$ to make clear which light
lepton is attributed to the $\PW$ boson and which to the Higgs boson. The $\Pe+\Pe\tauh$ final state is not considered because of the
lower acceptance and efficiency for electrons with respect to muons.
Throughout the paper neutrinos are omitted from the notation of the final states.

\section{The CMS detector}

The central feature of the CMS apparatus is a superconducting solenoid 6\unit{m} in
internal diameter, providing a magnetic field of 3.8\unit{T}. Within the solenoid
volume there are: a silicon pixel and strip tracker, a lead tungstate crystal
electromagnetic calorimeter (ECAL), and a brass and scintillator hadron calorimeter
(HCAL). Each of these is composed of a barrel and two endcap sections.
Forward hadron calorimeters extend
the pseudorapidity ($\eta$) coverage provided by the barrel and endcap detectors. Muons are
detected in gas-ionization chambers embedded in the steel flux-return yoke
outside the solenoid. Events are selected using a two-tiered trigger
system~\cite{Khachatryan:2016bia}.
A more detailed description of the CMS detector, together with a definition of
the coordinate system used and the relevant kinematic variables, can be found
in Ref.~\cite{Chatrchyan:2008zzk}.

\section{Simulated samples}

The signal samples with a Higgs boson produced in association with a $\PW$ or
a $\PZ$ boson ($\PW\PH$ or $\PZ\PH$) are generated at next-to-leading order
(NLO) in perturbative quantum chromodynamics (QCD) with the \POWHEG
2.0~\cite{Nason:2004rx,Frixione:2007vw, Alioli:2010xd, Alioli:2010xa, Alioli:2008tz}
generator extended with the MiNLO procedure~\cite{Luisoni:2013kna}.
The set of parton distribution functions (PDFs) is
NNPDF3.0~\cite{Ball:2011uy}. Because the analysis focuses on
measuring the $\PW\PH$ and $\PZ\PH$ processes, the $\ttbar\PH$ process is
included as a background. The contribution from Higgs boson events produced via gluon fusion or vector boson fusion is negligible
in this analysis. This is because the studied final states, when counting both 
leptonically and hadronically decaying $\PGt$ leptons, all include three or four charged lepton candidates.
The transverse momentum (\pt) distribution of
the Higgs boson in the {\POWHEG} simulations is tuned to match closely
the next-to-NLO (NNLO) plus next-to-next-to-leading-logarithmic prediction in the
\textsc{HRes 2.3} generator~\cite{deFlorian:2012mx,Grazzini:2013mca}.
The production cross sections and branching fractions for the SM Higgs
boson production and their corresponding uncertainties are taken from
Refs.~\cite{deFlorian:2016spz,Denner:2011mq,Ball:2011mu}.

The background samples of $\ttbar$, $\PW\PZ$, and $\cPq\cPq\to \PZ\PZ$ are generated at NLO with
$\POWHEG$, as are the $\PW\PH \to \PW\PW\PW$, $\PZ\PH \to \PZ\PW\PW$, and $\PH \to \PZ\PZ$ backgrounds.
The $\Pg\Pg \to \PZ\PZ$ process is generated at leading order (LO) with
\textsc{mcfm}~\cite{Campbell:2010ff}. The \MGvATNLO v2.3.3 generator is used for
triboson, $\ttbar \PW$, and $\ttbar \PZ$ production,
with the jet matching and merging scheme applied either at NLO with the FxFx algorithm~\cite{Frederix:2012ps}
or at LO with the MLM algorithm~\cite{Alwall:2007fs}.
The generators are interfaced with \PYTHIA 8.212~\cite{Sjostrand:2014zea}
to model the parton showering and fragmentation, as well as the decay of the $\PGt$ leptons.
The \PYTHIA parameters affecting the description of the underlying event are
set to the {CUETP8M1} tune~\cite{Khachatryan:2015pea}.

Generated events are processed through a simulation of the CMS detector based on
\GEANTfour~\cite{Agostinelli:2002hh}, and are reconstructed with the same algorithms
that are used for data.
The simulated samples include additional $\Pp\Pp$ interactions per bunch
crossing, referred to as pileup.
The effect of pileup is taken into account by generating concurrent minimum-bias
collision events. The simulated events are weighted such that the distribution of the number
of additional pileup interactions matches closely with data.  The pileup distribution in data is
estimated from the measured instantaneous luminosity for each bunch crossing and results in
an average of approximately 23 interactions per bunch crossing.

\section{Event reconstruction}
\label{sec:reconstruction}

The reconstruction of observed and simulated events relies on the
particle-flow (PF) algorithm~\cite{Sirunyan:2017ulk}. This algorithm combines
information from all subdetectors to identify
and reconstruct the particles emerging from $\Pp\Pp$ collisions:
charged hadrons, neutral hadrons, photons, muons, and electrons.
Combinations of these PF objects are used to reconstruct
higher-level objects such as
the missing transverse momentum (\ptvecmiss). The \ptvecmiss is defined as
the projection onto the plane perpendicular to the beam axis of the
negative vector sum of the momenta of all reconstructed particle-flow
objects in an event. Its magnitude is referred to as \ptmiss.
The primary $\Pp\Pp$ interaction vertex is taken to be the reconstructed vertex 
with the largest value of summed $\pt^2$ of jets and the associated \ptmiss, 
calculated from the tracks assigned to the vertex, where the jet finding algorithm 
is taken from Refs.~\cite{Cacciari:2008gp,Cacciari:2011ma} and the associated \ptvecmiss
is taken as the negative vector sum of the \pt of the jets.

Electrons are identified with a multivariate discriminant
combining several quantities describing the track quality,
the shape of the energy deposits in the ECAL,
and the compatibility of the measurements from the tracker and the
ECAL~\cite{Khachatryan:2015hwa}.
Muons are reconstructed by combining information from the inner
tracker and the muon systems, using two algorithms~\cite{Sirunyan:2018fpa}.
One matches tracks in the silicon tracker to hits in the muon detectors, while the other one performs a
track fit using hits in both the silicon tracker and the muon systems.
To reject nonprompt or misidentified leptons, a relative lepton isolation is defined as:
\begin{linenomath}
\begin{equation}
I^{\ell} \equiv \frac{\sum_\text{charged}  \PT + \max\left( 0, \sum_\text{neutral}  \PT
                 - \frac{1}{2} \sum_\text{charged, PU} \PT  \right )}{\PT^{\ell}}.
\label{eq:reconstruction_isolation}
\end{equation}
\end{linenomath}
In this expression, $\sum_\text{charged}  \PT$ is the scalar sum of the
transverse momenta of the charged particles originating from
the primary vertex and located in a cone of size
$\Delta R = \sqrt{\smash[b]{(\Delta \eta)^2 + (\Delta \phi)^2}} = 0.3$\,(0.4)
centered on the electron (muon) direction, where $\phi$ is the azimuthal angle in radians. The sum
$\sum_\text{neutral}  \PT$ represents
a similar quantity for neutral particles.
The contribution of photons and neutral hadrons originating from pileup
vertices is estimated from the scalar sum of the transverse
momenta of charged hadrons in the cone originating from pileup vertices,
$\sum_\text{charged, PU} \PT$. This sum is multiplied by a factor of
$1/2$, which corresponds approximately to the ratio of neutral to charged
hadron production in the hadronization process
of inelastic $\Pp\Pp$ collisions, as estimated from simulation.
The estimated contribution to $I^{\ell}$ from photons and neutral hadrons originating from the
primary vertex is required not to be negative, which is enforced
by the ``max`` notation in Eq.~\eqref{eq:reconstruction_isolation}.
The expression $\PT^{\ell}$ stands for the $\pt$ of the lepton.
Isolation requirements used in this analysis include
$I^{\Pe}<0.10$ and $I^{\Pgm}<0.15$ in the $\PW\PH$ channels.
In the $\PZ\PH$ channels, the isolation criteria are $I^{\Pe}<0.15$ ($I^{\Pgm}<0.15$) for electrons
(muons) associated to a $\tauh$ decay and $I^{\Pgm}<0.25$ for
muons associated to a $\PZ$ boson decay.

Jets are reconstructed with an anti-\kt clustering algorithm implemented
in the \FASTJET library~\cite{Cacciari:2011ma, Cacciari:fastjet2}.
It is based on the clustering of neutral and charged PF candidates with
a distance parameter of 0.4. Charged PF candidates not associated with
the primary vertex of the interaction are not considered when clustering.
The combined secondary vertex (CSVv2) algorithm is used to identify jets
that are likely to have originated from a bottom quark
(``$\cPqb$ jets'')~\cite{Sirunyan:2017ezt}. The algorithm
exploits the track-based lifetime information together with the secondary
vertices associated with the jet using a multivariate technique to produce a discriminator
for $\cPqb$ jet identification. A set of $\pt$-dependent correction
factors are applied as weights to simulated events to account for differences in the
b tagging efficiency between data and simulation~\cite{Sirunyan:2017ezt}. The working point chosen
in this analysis gives an identification efficiency for genuine $\cPqb$ jets of about 70\%
and a misidentification probability for light flavor or gluon jets of about 1\%.
All events with a \cPqb-tagged jet are discarded from this analysis. This selection
requirement suppresses the contributions of $\ttbar$, $\ttbar+\PW$, and $\ttbar+\PZ$
with minimal impact to the signal selection efficiency.

Hadronically decaying $\PGt$ leptons are reconstructed with the hadron-plus-strips (HPS)
algorithm~\cite{Khachatryan:2015dfa, CMS-PAS-TAU-16-002}, which is
seeded from anti-\kt jets. The HPS algorithm reconstructs $\tauh$
candidates on the basis of the number of tracks and on the number of ECAL
strips with an energy deposit in the $\eta$-$\phi$ plane, in the 1-prong,
1-$\text{prong}+\PGpz$, and 3-prong decay modes. A
multivariate analysis (MVA) discriminator~\cite{Hocker:2007ht}, including isolation
and lifetime information, is used to reduce the rate for quark- and gluon-initiated jets
to be identified as $\tauh$ candidates. The three working points used in this analysis
have efficiencies of about 55, 60, and 65\% for genuine $\tauh$,
with about 1.0, 1.5, and 2.5\% misidentification rates for quark- and gluon-initiated jets,
within a $\pt$ range typical of a $\tauh$ originating from a $\PZ$ boson. The first
working point is used in the $\ell+\tauh\tauh$ channels of $\PW\PH$ for the $\tauh$ that has
the same charge as the electron or muon, while the third working point is used for the
$\tauh$ that has the opposite charge. The second working point is used in the $\PW\PH$
channels with exactly one $\tauh$.
The third working point is used for all $\tauh$ in the $\PZ\PH$ channels. 
Electrons misidentified as $\tauh$ candidates are suppressed using a second MVA discriminator that
includes tracker and calorimeter information~\cite{CMS-PAS-TAU-16-002}.
Muons misidentified as $\tauh$ candidates are suppressed using additional
cut-based criteria requiring energy and momentum consistency between the measurements in the
tracker and the calorimeters, and requiring no more than one segment in the muon detectors~\cite{Khachatryan:2015dfa}.
The working points of these discriminators are specific to each decay channel. The $\tauh$ energy in simulation is corrected
for each decay mode on the basis of a measurement of the $\tauh$ energy scale in $\PZ\to\PGt\PGt$ events.
The rate and the energy of electrons and muons misidentified as $\tauh$
candidates are also corrected in simulation on the basis of a ``tag-and-probe''
measurement~\cite{CMS:2011aa} in $\PZ\to\ell\ell$ events.

In all final states, the visible mass of the Higgs boson candidate, $\mvis$, can be used to separate
the $\PH\to \PGt \PGt$ signal events
from the large irreducible contribution of $\PZ \to \PGt \PGt$ events.
However, the neutrinos from the $\PGt$ lepton decays carry a large fraction of
the $\PGt$ lepton energy and reduce the discriminating power of this variable.
The \textsc{svfit} algorithm~\cite{Bianchini:2014vza} combines \ptvecmiss
with the four-vector momenta of both $\PGt$ candidates to estimate the mass of the parent boson, denoted as $\mtt$. The resolution
of $\mtt$ is about 20\%. The
$\mtt$ variable is used for the $\PZ\PH$ channels, while $\mvis$ is used for the
$\PW\PH$ channels because the \textsc{svfit} algorithm cannot account for the
additional \ptvecmiss from the $\PW$ boson decay.

\section{Event selection}
\label{sec:selection}

Events for the $\PW\PH$ and $\PZ\PH$ production channels are selected using single- or
double-lepton triggers targeting leptonic decays of the $\PW$ and $\PZ$ bosons.
The trigger and offline selection requirements for
all possible decay modes are presented in Table~\ref{tab:inclusive_selection}.
Leptons selected by the trigger must be matched to those selected
in the analysis.
The light leptons (electrons and muons) in the events are required to be separated from each
other by $\Delta R > 0.3$, while the $\tauh$ candidates must be separated from each other and
from the other leptons by $\Delta R > 0.5$. The resulting event samples are made mutually
exclusive by discarding events that have additional identified
and isolated electrons or muons.

\begin{table*}[htbp]
\centering
\topcaption{Kinematic selection requirements for $\PW\PH$ and $\PZ\PH$ events.
The trigger requirement is defined by a combination of trigger candidates with
\pt over a given threshold (in \GeV), indicated inside parentheses. The
$\abs\eta$ thresholds come from trigger and object reconstruction constraints.
$\PZ\PH$ events are selected with either a lower \pt threshold double lepton trigger
or a higher \pt threshold single lepton trigger.
\label{tab:inclusive_selection}
}
\cmsTable{
    \begin{tabular}{llll}
         \multicolumn{4}{c}{$\PW\PH$ selection}                 \\
         \multicolumn{4}{c}{$\tauh$ baseline requirements: $\pt^{\tauh}>20\GeV$, $\abs{\eta^{\tauh}}<2.3$}   \\
         \multicolumn{4}{c}{$\Pe$ baseline requirements: $\pt^\Pe>15\GeV$, $\abs{\eta^\Pe}<2.5$, $\Pe$ ID 80\% efficiency, $I^\Pe<0.10$}   \\
         \multicolumn{4}{c}{$\Pgm$ baseline requirements: $\pt^\Pgm>15\GeV$, $\abs{\eta^\Pgm}<2.4$, $\Pgm$ ID $> 99\%$ efficiency, $I^\Pgm<0.15$ }   \\
    \hline
      Channel           &         Trigger ($\pt (\GeVns)/\abs\eta$)         & Light lepton selection  & $\tauh$ selection  \\
    \hline
     $\Pe+\Pgm\tauh$/$\Pgm+\Pe\tauh$      &  $\Pe (25/2.1)$ or $\Pgm (22/2.1)$                     &     $\pt^\Pe>26\GeV$ or $\pt^\Pgm>23\GeV$                          &  $\tauh$ isolation 60\% eff.  \\
     $\Pgm+\Pgm\tauh$     &  $\Pgm (22/2.1)$                    &     $\pt^\Pgm>23\GeV$                         &  $\tauh$ isolation 60\% eff.  \\
     $\Pe+\tauh\tauh$     &  $\Pe (25/2.1)$                     &     $\pt^\Pe>26\GeV$                          &  $\tauh$ isolation 55 or 65\% eff.  \\
     $\Pgm+\tauh\tauh$    &  $\Pgm (22/2.1)$                    &     $\pt^\Pgm>23\GeV$                         &  $\tauh$ isolation 55 or 65\% eff.  \\

    \\
         \multicolumn{4}{c}{$\PZ\PH$ selection}                 \\
         \multicolumn{4}{c}{$\PZ$ boson reconstructed from opposite charge, same-flavor light leptons, $60 < m_{\ell\ell} < 120\GeV$}  \\
         \multicolumn{4}{c}{$\tauh$ baseline requirements: $\pt^{\tauh}>20\GeV$, $\abs{\eta^{\tauh}}<2.3$, $\tauh$ isolation 65\% efficiency}   \\
         \multicolumn{4}{c}{$\Pe$ baseline requirements: $\pt^\Pe>10\GeV$, $\abs{\eta^\Pe}<2.5$, $\Pe$ ID 90\% efficiency}   \\
         \multicolumn{4}{c}{$\Pgm$ baseline requirements: $\pt^\Pgm>10\GeV$, $\abs{\eta^\Pgm}<2.4$, $\Pgm$ ID $> 99\%$ efficiency, $I^\Pgm<0.25$ }   \\
    \hline
      Channel           &         Trigger ($\pt (\GeVns)/\abs\eta$)         & $\PZ \to \ell\ell$ lepton selection   & $\htt$ lepton selection  \\
    \hline
      $\Pe\Pe+\Pgm\tauh$     &                                    &                                    &  $I^\Pgm<0.15$       \\
      $\Pe\Pe+\Pe\tauh$      & $\left[\Pe_{1}(23/2.5)\,\&\,\Pe_{2}(12/2.5)\right]$  &  $\left[\pt^{\Pe_{1}}>24\GeV~\&\,\pt^{\Pe_{2}}>13\GeV\right]$     &  $\Pe$ ID 80\% eff., $I^\Pe<0.15$ \\
      $\Pe\Pe+\tauh\tauh$    & or $\Pe_{1}(27/2.5)$                   &  or $\pt^{\Pe_{1}}>28\GeV$                   &  baseline selection listed above      \\
      $\Pe\Pe+\Pe\Pgm$       &                                    &                                    &  $\Pe$ ID 80\% eff., $I^\Pe<0.15$, $I^\Pgm<0.15$ \\
    \hline
      $\Pgm\Pgm+\Pgm\tauh$   &                                    &                                    &  $I^\Pgm<0.15$       \\
      $\Pgm\Pgm+\Pe\tauh$    &  $\left[\Pgm_{1}(17/2.4)\,\&\,\Pgm_{2}(8/2.4)\right]$ &  $\left[\pt^{\Pgm_{1}}>18\GeV~\&\,\pt^{\Pgm_{2}}>10\GeV\right]$   &  $\Pe$ ID 80\% eff., $I^\Pe<0.15$ \\
      $\Pgm\Pgm+\tauh\tauh$  &   or $\Pgm_{1}(24/2.4)$                &  or $\pt^{\Pgm_{1}}>25\GeV$                  &  baseline selection listed above      \\
      $\Pgm\Pgm+\Pe\Pgm$     &                                    &                                    &  $\Pe$ ID 80\% eff., $I^\Pe<0.15$, $I^\Pgm<0.15$ \\

    \end{tabular}
}
\end{table*}

In the $\Pe+\Pgm\tauh$/$\Pgm+\Pe\tauh$ and $\Pgm+\Pgm\tauh$ final states of the $\PW\PH$ channel,
the two light leptons are required to have the same charge to reduce the $\ttbar$
and $\zJets$ backgrounds where one or more jets is misidentified as a $\tauh$
candidate. The highest $\pt$
light lepton is considered as coming from the $\PW$ boson.
The Higgs boson candidate is formed from the $\tauh$ candidate, which
must have opposite charge to the light leptons, and the subleading light lepton.
The correct pairing is achieved in about 75\% of events, according to simulation. The leading light lepton is required
to pass a single-lepton trigger and to have a $\pt$ that is 1\GeV above the online
threshold, whereas the subleading light lepton must have $\pt>15\GeV$, as determined from
optimizing for signal sensitivity. 
In $\PW\PH$ associated production, the Higgs and $\PW$ bosons are dominantly produced 
back-to-back in $\phi$, and may have a longitudinal Lorentz boost that makes them close in $\eta$.
There is an increased background of misidentified jets at high $\eta$ because of the 
decreased detector performance in the endcaps.
Considering these characteristics, selection criteria based on three variables have been found to
improve the signal sensitivity in both the $\Pe+\Pgm\tauh$/$\Pgm+\Pe\tauh$ and $\Pgm+\Pgm\tauh$ final states:
\begin{itemize}
\item $\LT>100\GeV$, where $\LT$ is the scalar sum of \pt of the light leptons and the $\tauh$ candidate;
\item $\abs{\Delta\phi(\ell_1,\PH)}>2.0$, where $\ell_1$ is the leading light lepton, and
$\PH$ is the system formed by the subleading light lepton and the $\tauh$ candidate;
\item $\abs{\Delta\eta(\ell_1,\PH)}<2.0$.
\end{itemize}

In the $\Pe+\tauh\tauh$ and $\Pgm+\tauh\tauh$ final states of the $\PW\PH$ channel,
the $\tauh$ candidates are required to have opposite charge. The $\tauh$ candidate that has the same charge as the light lepton must
have $\pt > 35\GeV$, while the other one is required to have $\pt > 20\GeV$. This requirement is driven
by the fact that the $\tauh$ candidate with the same charge as the light lepton
is often a jet misidentified as a $\tauh$ from the SM background, and
the jet misidentification
rate strongly decreases as $\pt$ increases. Selection criteria based on three variables
have been found to improve the results in the $\Pe+\tauh\tauh$ and $\Pgm+\tauh\tauh$ final states:
\begin{itemize}
\item $\LT>130\GeV$, where $\LT$ is the scalar sum of \pt of the light lepton and $\tauh$ candidates;
\item $\abs{\vST}<70\GeV$, where $\vST$ is the vector sum of \pt of the light lepton, $\tauh$ candidates, and $\ptvecmiss$;
\item $\abs{\Delta\eta(\tauh,\tauh)}<2.0$.
\end{itemize}

In the $\PZ\PH$ final states, the $\PZ$ boson is reconstructed from the opposite charge, same-flavor
light lepton combination that has a mass closest to the $\PZ$ boson mass. 
Different identification and isolation selections are applied to the light leptons associated to 
the $\PZ$ boson compared with those associated to the Higgs boson. The selections are
looser for those associated with the $\PZ$ boson to increase the signal acceptance,
while tighter selections are applied to the light leptons assigned to the Higgs boson to
decrease the background contributions from $\zJets$ and other reducible
backgrounds. 
The leptons assigned to the Higgs boson are required to have opposite charge.
The specific selections detailed in Table~\ref{tab:inclusive_selection},
including those chosen for the $\tauh$ candidates,
were optimized to obtain the best expected signal sensitivity.

Candidates for associated $\PZ\PH$ production are also categorized depending on the value of $\LTH$,
defined as the scalar sum of $\pt$ of the visible decay
products of the Higgs boson.
The large Higgs boson mass causes the decay products to have relatively high $\pt$
compared to the jets misidentified as leptons from the $\zJets$ background process,
which leads to a higher signal purity in the category with high $\LTH$. The thresholds to separate the high $\LTH$ and low $\LTH$
regions are optimized to maximize the expected signal sensitivity for each $\htt$ final state. The threshold is equal to 50\GeV in the $\ell\ell+\Pe\Pgm$
final states, 60\GeV in the $\ell\ell+\Pe\tauh$ and $\ell\ell+\Pgm\tauh$ final states, and 75\GeV in the $\ell\ell+\tauh\tauh$ final state.

\section{Background estimation}
\label{sec:background_estimation}

The irreducible backgrounds ($\PZ\PZ$, $\ttbar\PZ$, $\PW\PW\PZ$, $\PW\PZ\PZ$,
$\PZ\PZ\PZ$, as well as $\PW\PZ$ and $\ttbar \PW$ in the $\PW\PH$ channels) are
estimated from simulation and scaled by their theoretical cross sections at the
highest order available. Inclusive Higgs
boson decays to $\PW$ or $\PZ$ boson pairs and the $\ttbar\PH$ associated production background
processes are also estimated from simulation.

The reducible backgrounds, which have at least one jet misidentified as an electron,
muon, or $\tauh$ candidate, are estimated from data. The dominant reducible background contributions come
from $\ttbar$ and $\zJets$ in the $\PW\PH$ channels and from $\ttbar$, $\zJets$,
and $\PW\PZ+\text{jets}$ in the $\PZ\PH$ channels. 
Misidentification rates are estimated in control samples that specifically measure the rate at which jets pass 
the identification criteria used for each $\Pgt$ candidate (electrons, muons, or $\tauh$). The 
misidentification rates are then applied to reweight events with $\Pgt$ candidates failing the 
identification criteria but passing all other signal region selections. These reweighted events estimate the contribution 
from processes with jets misidentified as $\Pgt$ candidates in the signal region.

In the $\PW\PH$ analysis, the misidentification rate of jets as $\Pgt$ candidates is measured
in $\zJets$ events. After reconstructing the $\PZ\to\Pe\Pe$ decay, the jet-to-muon
misidentification rate is estimated as a function of the lepton $\pt$ by applying the
lepton identification algorithm to any additional jet in the event. Similarly,
$(\PZ\to\Pgm\Pgm)+\text{jets}$ events are used to estimate the jet-to-electron and
jet-to-$\tauh$ misidentification rates. Events where the $\Pgt$ candidates arise from genuine leptons,
primarily from the $\PW\PZ$ process, are estimated from simulation and subtracted
from the data so that the misidentification rates are measured for jets only.
The rates are measured in bins of lepton $\pt$, and are
separated by the reconstructed decay mode of the $\tauh$ candidates.

In the $\Pe+\Pgm\tauh$/$\Pgm+\Pe\tauh$ and $\Pgm+\Pgm\tauh$ final states,
events that do not pass the identification
conditions of either the subleading light lepton or the $\tauh$ are reweighted to estimate the reducible background contribution in the signal region. In particular,
events with exactly one object failing the identification criteria receive a weight $f/(1-f)$,
where $f$ is the misidentification rate for the particular type of object. Events with both objects
failing the identification criteria receive a weight $-f_1f_2/[(1-f_1)(1-f_2)]$, where the negative
sign removes the double counting of events with two jets.
This method estimates the number of events for which the subleading light lepton or the $\tauh$
candidate corresponds to a jet. Such events are therefore removed from simulated
samples to avoid double counting. However,
events that have a jet misidentified as the leading lepton, but two genuine leptons
for the subleading lepton and the $\tauh$, are not taken into account with the misidentification rate method and
are therefore estimated from simulation. These events mostly arise from $\ttbar$ and $\zJets$ processes,
and account for less than 10\% of the total expected background in the signal region.
In the $\Pe+\tauh\tauh$ and $\Pgm+\tauh\tauh$ final states of the $\PW\PH$ channels,
the method is essentially the same, except that the misidentification rate functions are applied
only to events where the $\tauh$ candidate that has the same charge as the light lepton fails the identification
criteria.

In the $\PZ\PH$ analysis, a very similar method is used
to estimate the contribution of jets misidentified as electrons, muons, or $\tauh$
candidates in the signal region. The misidentification rates are measured in a region
with an opposite-charge same-flavor lepton pair compatible with a $\PZ$ boson, and two additional objects. This
region is dominated by $\zJets$ events with a small contribution from
$\ttbar$ events. In a procedure identical to that of the $\PW\PH$ final states,
the contribution from genuine leptons is estimated from simulation and is subtracted, and the rates are measured
in bins of lepton $\pt$ and are split between reconstructed decay modes for
the $\tauh$ candidates.
In the $\PZ\PH$ analysis, events that pass the full
signal region selection with the exception that either or both of the $\Pgt$ candidates associated
to the Higgs boson fail the identification criteria are weighted as a function of the misidentification
rates.
To avoid double counting, events with both $\Pgt$ candidates failing the selection criteria have their weight
subtracted from the events that have only a single object failing.
This misidentification rate method is used to estimate only the yield of the reducible
backgrounds. The $\mtt$ distribution of the reducible background contribution is taken from
data in a region with negligible signal and irreducible background contribution, defined similarly to the signal region
but with same charge $\Pgt$ candidates passing relaxed identification and isolation criteria.

\section{Systematic uncertainties}
\label{sec:systematics}

The overall uncertainty in the $\tauh$ identification efficiency for genuine $\tauh$
leptons is 5\%~\cite{CMS-PAS-TAU-16-002}, which has been measured with a tag-and-probe method in $\PZ\to\PGt\PGt$ events.
An uncertainty of 1.2\% in the visible energy of genuine $\tauh$ leptons
affects both the shape and yield of the final mass distributions for the signals and backgrounds. It is
uncorrelated among the 1-prong, 1-$\textrm{prong}+\PGpz$, and 3-prong decay modes.

The uncertainties in the electron and muon identification, isolation, and trigger
efficiencies lead to a rate uncertainty of 2\% for both electrons and muons.
The uncertainty in the electron energy, which amounts to 2.5\% in the endcaps
and 1\% in the barrel, affects both the shape and yield of the final mass distributions.
In all channels, the effect of the uncertainty in
the muon energy is negligible.

The rate uncertainty related to discarding events with a \cPqb-tagged jet is
4.5\% for processes with heavy-flavor jets, and 0.15\% for processes with light-flavor jets.

Theoretical uncertainties associated with finite-order perturbative calculations, and with the
choice of the PDF set, are taken into account for the $\PZ\PZ$ and $\PW\PZ$
background processes. The theoretical uncertainties are evaluated
by varying renormalization and factorization scales by factors of
0.5 and 2.0, independently. The process leads to yield uncertainties
of $^{+3.2\%}_{-4.2\%}$ for the $\Pq\Pq\to \PZ\PZ$ process, and $\pm 3.2\%$
for the $\PW\PZ$ process. The uncertainty from the PDF set is determined to be
$^{+3.1\%}_{-4.2\%}$ for
the $\Pq\Pq\to \PZ\PZ$ process, and $\pm 4.5\%$ for the $\PW\PZ$ process.
In addition, a 10\% uncertainty in the NLO $K$ factor used for the $\Pg\Pg \to
\PZ\PZ$ prediction is used~\cite{Sirunyan:2017exp}. The uncertainties in the cross section of the
rare $\ttbar\PW$ and $\ttbar\PZ$ processes amount to 25\%~\cite{Sirunyan:2017uzs}.

The rate and acceptance uncertainties for the signal processes related to the
theoretical calculations arise from uncertainties in the PDFs, variations of
the QCD renormalization and factorization scales, and uncertainties in the
modeling of parton showers. The magnitude of the rate uncertainty is estimated
from simulation and depends on the production process.
The inclusive uncertainties related to the PDFs amount to 1.9 and 1.6\%,
respectively, for the $\PW\PH$ and $\PZ\PH$ production modes~\cite{deFlorian:2016spz}. The
corresponding uncertainty for the variation of the renormalization and
factorization scales is 0.7 and 3.8\%, respectively~\cite{deFlorian:2016spz}.

The reducible backgrounds are estimated by using the measured rates for jets to be
misidentified as electron, muon, or $\tauh$ candidates.
In the $\PW\PH$ channels, an
uncertainty arises from potentially different misidentification rates
in $\zJets$ events, where the rates are measured, and in $\PW+\text{jets}$ or
$\ttbar$ events, which constitute a large fraction of the reducible background
in the signal region. This leads to a 20\% yield uncertainty for the reducible
background in each final state of the $\PW\PH$ analysis. This uncertainty
also covers the measured differences in observed versus predicted reducible background
yields in multiple dedicated control regions.

In the $\PZ\PH$ final
states a similar uncertainty is applied based on potential differences
between the region where the misidentification rates are measured and the region where they are applied.
These uncertainties are based on the results of closure tests comparing the differences in observed
versus predicted reducible background yields.  The uncertainty is taken to be the largest difference
between simulation-based and data-based closure tests.
The yield uncertainties are 50\%
in the $\ell\ell+\Pe\tauh$ final states, 25\% in $\ell\ell+\Pgm\tauh$, 40\% in $\ell\ell+\tauh\tauh$,
and 100\% in $\ell\ell+\Pe\Pgm$.
The large uncertainty in the $\ell\ell+\Pe\Pgm$
final states results from the very low expected reducible background yields, which
makes the closure tests susceptible to large statistical fluctuations.

The misidentification rates of jets as $\Pgt$ candidates are
measured in different bins of lepton $\pt$, separately for the three
reconstructed decay modes for the $\tauh$ candidate. In the $\PW\PH$ channels, where
the mass distribution for the reducible background is taken from the misidentification rate
method, the statistical uncertainty in every
bin is considered as an independent uncertainty and is propagated to the mass
distributions and to the yields of the reducible background estimate. In contrast, in the $\PZ\PH$ channels,
the mass distribution of the reducible background is estimated from data in
a region where the $\Pgt$ candidates have the same charge and pass relaxed isolation conditions.
Therefore, the statistical uncertainties
in the misidentification rates do not have an impact on the shape of the mass
distribution in this channel. Additionally, their impact on the reducible background yields
is subleading compared to the closure-based uncertainties.
In both the $\PW\PH$ and $\PZ\PH$ channels, an additional
uncertainty in the misidentification rates arising from the subtraction of
prompt leptons estimated from simulation is taken into account and propagated to
the reducible background mass distributions.

The \ptvecmiss scale uncertainties~\cite{CMS-PAS-JME-16-004}, which are computed
event-by-event, affect the normalization of various processes through the event
selection, as well as their distributions through the propagation of these
uncertainties to the di-$\PGt$ mass $\mtt$ in the $\PZ\PH$ channels. The
\ptvecmiss scale uncertainties arising from unclustered energy deposits in the
detector come from four independent sources related to the tracker, ECAL, HCAL,
and forward calorimeters. Additionally, \ptvecmiss scale
uncertainties related to the uncertainties in the jet energy measurement,
which affect the \ptvecmiss calculation, are taken into
account.

Uncertainties related to the finite number of simulated events, or to the
limited number of events in data control regions, are taken into account. They
are considered for all bins of the distributions used to extract the results.
They are uncorrelated across different samples, and across bins of a single distribution.
Finally, the uncertainty in the integrated luminosity amounts to 2.5\%~\cite{CMS-PAS-LUM-17-001}.
The systematic uncertainties considered in the analysis are summarized in Table~\ref{tab:uncertainties}.

\begin{table*}[!ht]
\centering
\topcaption{Sources of systematic uncertainty. The sign $\dagger$ marks the uncertainties that
are both shape- and rate-based. Uncertainties that affect only the normalizations have no marker.
For the shape and normalization uncertainties, the magnitude column lists the range of the associated change in normalization,
which varies by process and final state.
The last column specifies the processes affected by each source of uncertainty.
}
\newcolumntype{x}{D{,}{\text{--}}{2.2}}
\begin{tabular}{lll}
Source of uncertainty & Magnitude & Process \\
\hline
 $\tauh$ ID \& isolation                            & 5\%  & All simulations \\
 $\tauh$ energy$^{\dagger}$ (1.2\% energy shift)                   & 0.1--1.9\% & All simulations \\
 $\Pe$ ID \& isolation \& trigger                     & 2\% & All simulations \\
 $\Pe$ energy$^{\dagger}$ (1--2.5\% energy shift)                       & 0.3--1.4\% & All simulations\\
 $\Pgm$ ID \& isolation \& trigger                  & 2\% & All simulations \\
 $\cPqb$ veto                                             & 0.15--4.50\% & All simulations\\
 Diboson theoretical uncertainty                    & 5\% & $\PW\PZ$, $\PZ\PZ$ \\
 $\Pg\Pg \to \PZ\PZ$ NLO $K$ factor                 & 10\% & $\Pg\Pg \to \PZ\PZ$\\
 $\ttbar+\PW$/$\PZ$ theoretical uncertainty         & 25\% & $\ttbar+\PW$/$\PZ$ \\
 Signal theoretical uncertainty                     & Up to 4\%, see text & Signal \\
 Reducible background uncertainties:                &   & Reducible bkg. \\
 $\quad\PW\PH$ statistical error propagation$^{\dagger}$         & 1--2\% &  \\
 $\quad\PW\PH$ prompt lepton normalization$^{\dagger}$           & 2.6\% in $\Pe+\Pgm\tauh$/$\Pgm+\Pe\tauh$, 4\% in $\Pgm+\Pgm\tauh$ & \\
 $\quad\PZ\PH$ prompt lepton normalization$^{\dagger}$           & 20\% in $\ell\ell+\Pe\Pgm$, $<$1\% elsewhere & \\
 $\quad\PW\PH$ normalization                     & 20\% & \\
 $\quad\PZ\PH$ normalization                     & 25--100\% &  \\
 $\ptvecmiss$ energy$^{\dagger}$              & Up to 1.5\% in $\PW\PH$, $<$1\% in $\PZ\PH$ & All simulations \\
 Limited number of events                           & Stat. uncertainty per bin & All \\
 Integrated luminosity                              & 2.5\% & All simulations \\
\end{tabular}
\label{tab:uncertainties}
\end{table*}

\section{Results}
\label{sec:results}

\begin{figure}[h!t]
\centering
  \includegraphics[width=0.45\textwidth]{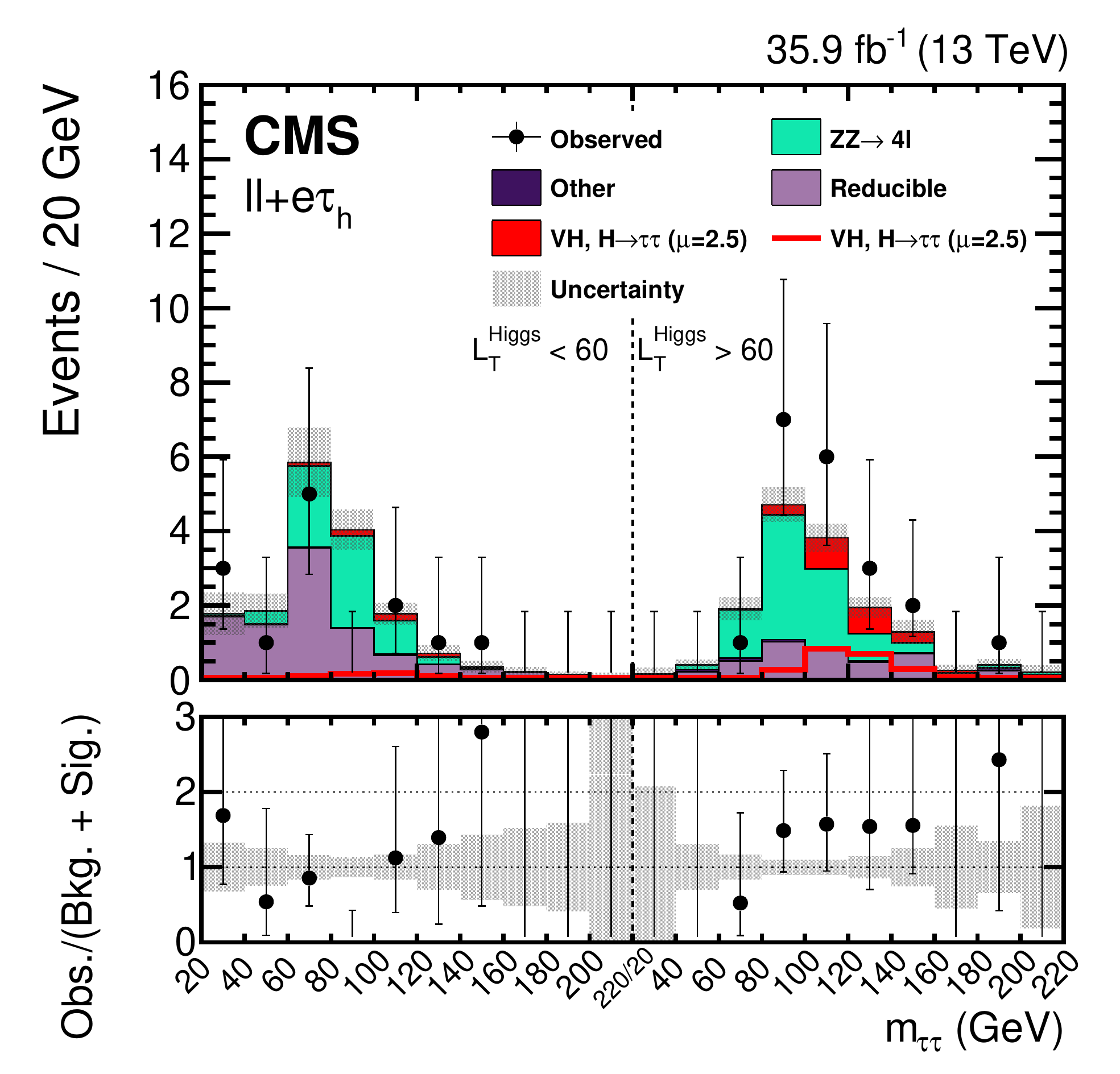}
  \includegraphics[width=0.45\textwidth]{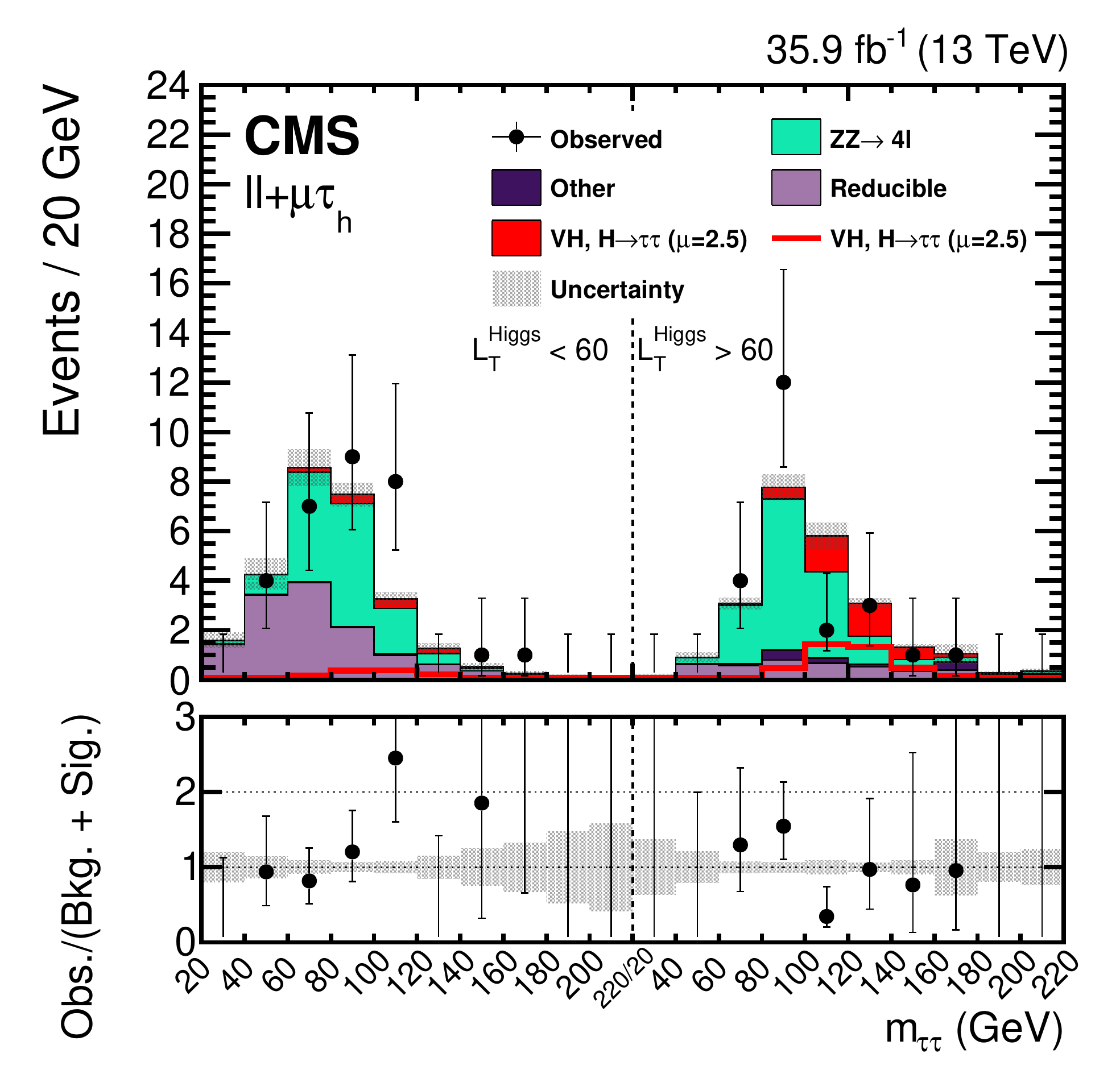}
  \includegraphics[width=0.45\textwidth]{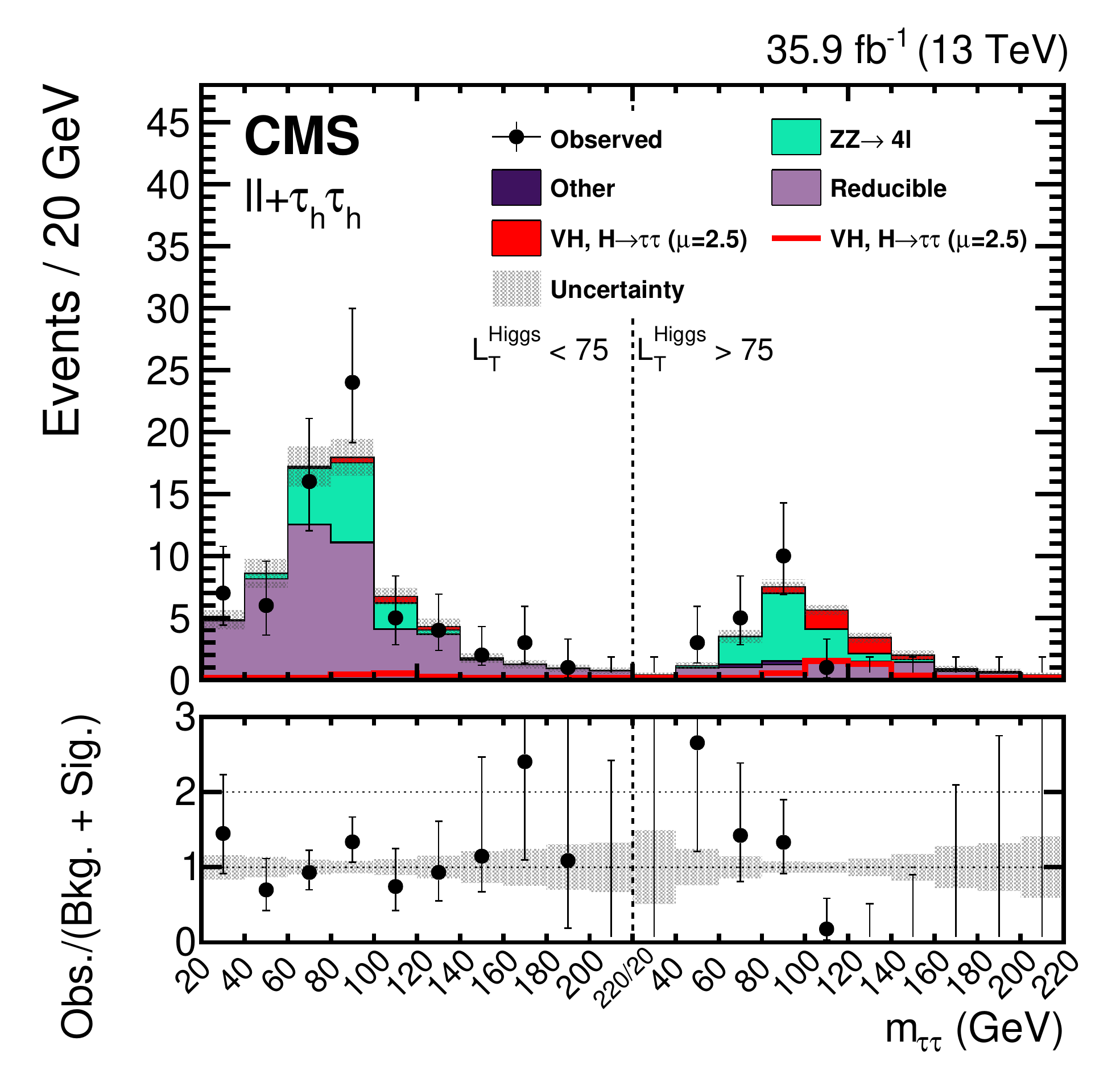}
  \includegraphics[width=0.45\textwidth]{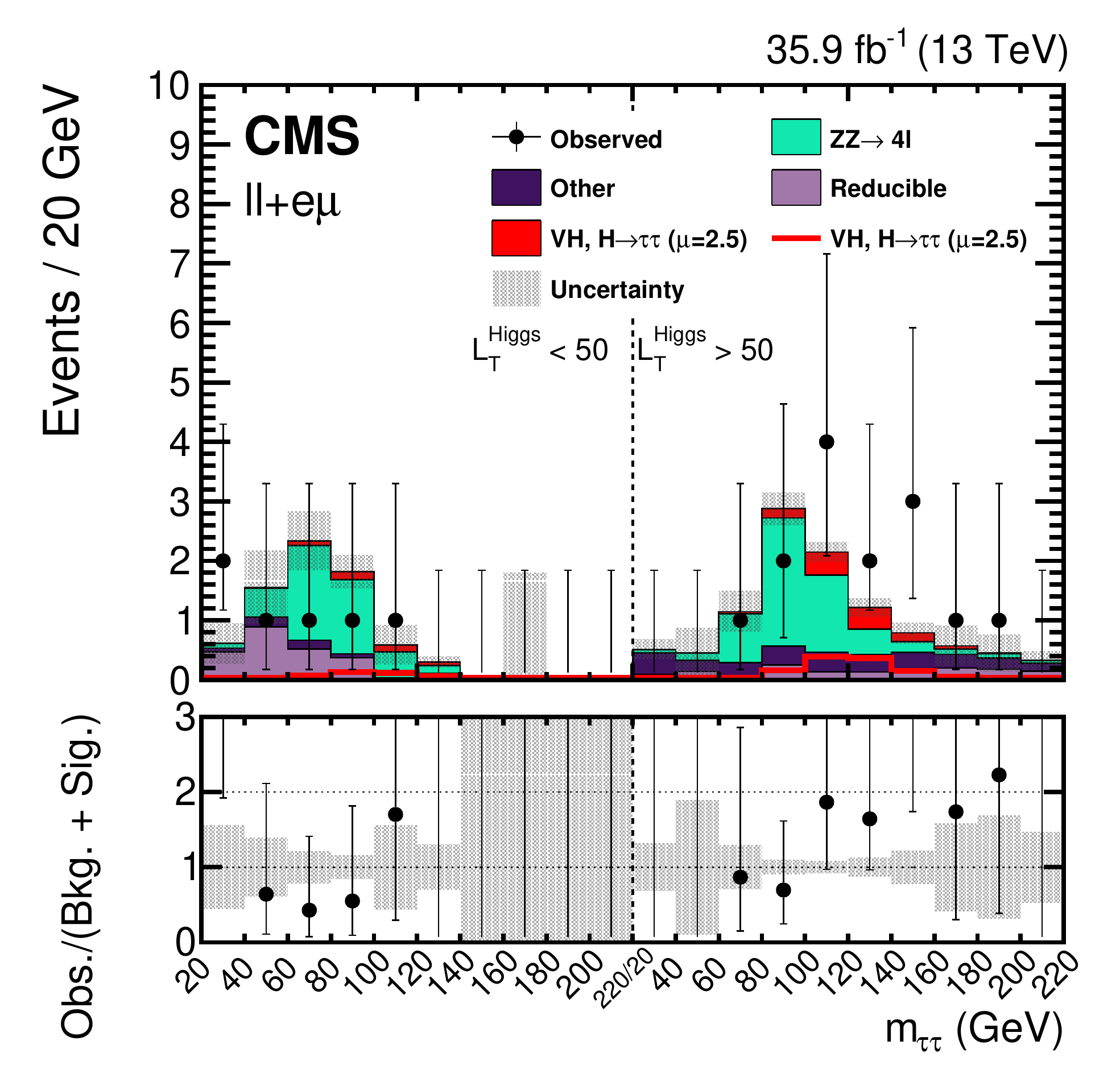}
 \caption{The post-fit $\mtt$ distributions used to extract the signal shown
  for (upper left) $\ell\ell+\Pe\tauh$, (upper right) $\ell\ell+\Pgm\tauh$,
  (lower left) $\ell\ell+\tauh\tauh$, and (lower right) $\ell\ell+\Pe\Pgm$.
  The uncertainties include both statistical and systematic components.
  The left half of each distribution is the low $\LTH$ region,
  while the right half of each distribution is the high $\LTH$ region.
  The $\PW\PH$ and $\PZ\PH$, $\htt$ signal processes are summed
  together and shown as $\VH$, $\htt$ with a best fit $\mu = 2.5$.
  $\VH$, $\htt$ is shown both as a stacked filled histogram and an
  open overlaid histogram.
  The contribution from
  ``Other'' includes events from triboson, $\ttbar+\PW$/$\PZ$, $\ttbar\PH$ production,
  and all production modes leading to $\hww$ and $\hzz$ decays.
  In these distributions the $\PZ\PH$, $\htt$ process contributes more than
  99\% of the total of $\VH$, $\htt$.
 }
 \label{fig:zh_results_svFitLLXX}
\end{figure}

\begin{figure}[h!t]
\centering
  \includegraphics[width=0.65\textwidth]{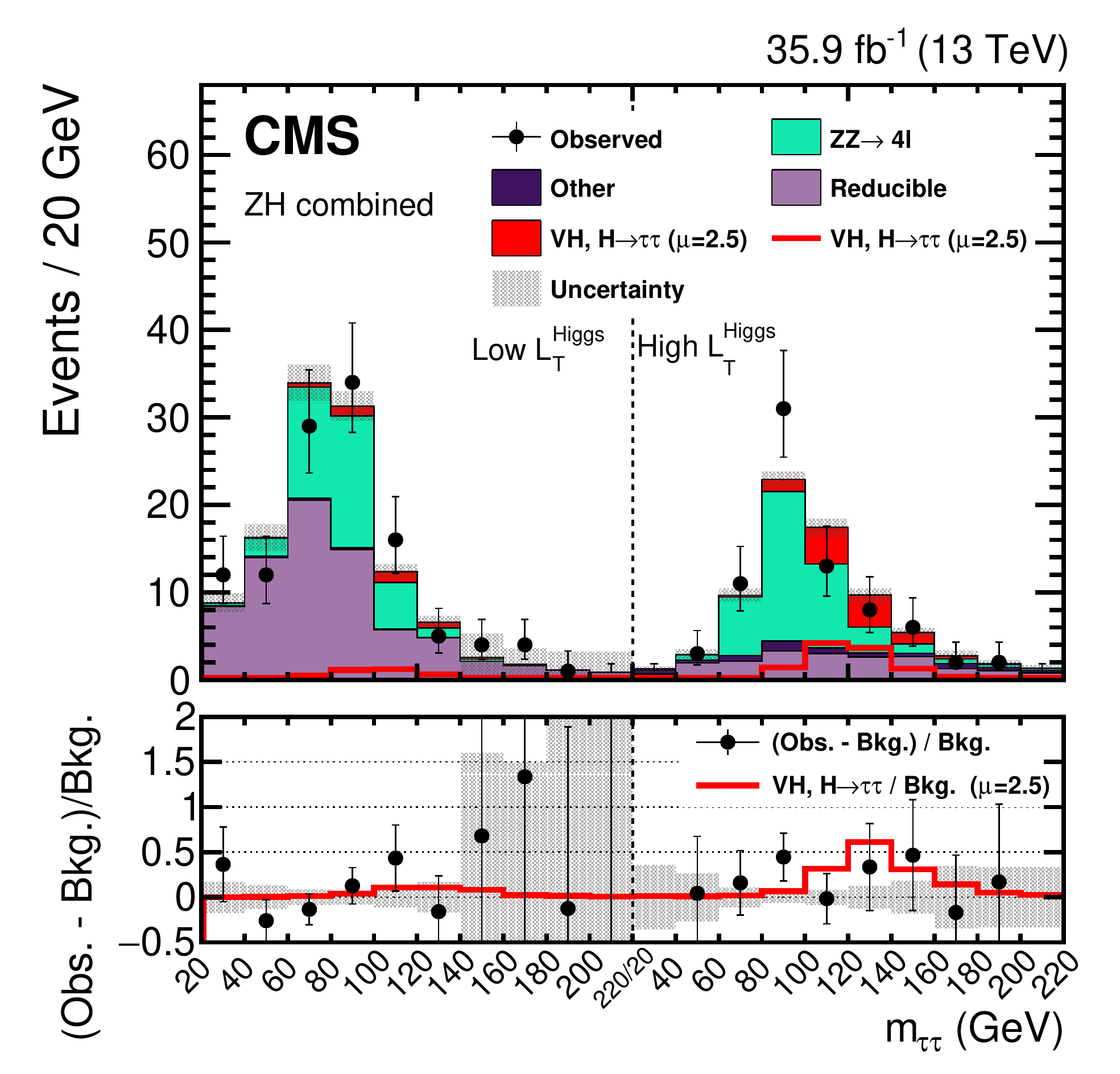}
 \caption{The post-fit $\mtt$ distributions used to extract the signal, shown
  for all 8 $\PZ\PH$ channels combined.
  The uncertainties include both statistical and systematic components.
  The left half of the distribution is the low $\LTH$ region,
  while the right half corresponds to the high $\LTH$ region.
  The definitions of the $\LTH$ regions in this distribution
  are the same as those used in Fig.~\ref{fig:zh_results_svFitLLXX} and are
  final state dependent.
  The $\PW\PH$ and $\PZ\PH$, $\htt$ signal processes are summed
  together and shown as $\VH$, $\htt$ with a best fit $\mu = 2.5$.
  $\VH$, $\htt$ is shown both as a stacked filled histogram and an
  open overlaid histogram.
  The contribution from
  ``Other'' includes events from triboson, $\ttbar+\PW$/$\PZ$, $\ttbar\PH$ production,
  and all production modes leading to $\hww$ and $\hzz$ decays.
  In this distribution the $\PZ\PH$, $\htt$ process contributes more than
  99\% of the total of $\VH$, $\htt$.
 }
 \label{fig:zh_results_svFitAll_mod}
\end{figure}

The results of the analysis are extracted with a global maximum likelihood fit
based on the reconstructed Higgs boson mass distributions in the eight $\PZ\PH$ and
four $\PW\PH$ signal regions. In the $\PZ\PH$ channels, the $\mtt$ distribution is used.
The $\mtt$ distributions are shown in Fig.~\ref{fig:zh_results_svFitLLXX} for each of the four $\htt$ final states,
and in Fig.~\ref{fig:zh_results_svFitAll_mod} for all eight $\PZ\PH$ channels
combined together. The low $\LTH$ and
high $\LTH$ regions are plotted side-by-side.
The eight $\PZ\PH$ channels are each fit as separate distributions in the global
fit; combining them together is for visualization purposes only.
The $\PW\PH$ and $\PZ\PH$ signal yields correspond to their best fit
signal strength value of 2.5.
The distributions are shown after the fit and include both
statistical and systematic uncertainties.
The signal and background predicted yields, as well as the number of observed events, are given
for each of the four $\htt$ final states of the $\PZ\PH$ channel in Table~\ref{tab:sb_zh}.

\begin{figure}[h!t]
\centering
  \includegraphics[width=0.45\textwidth]{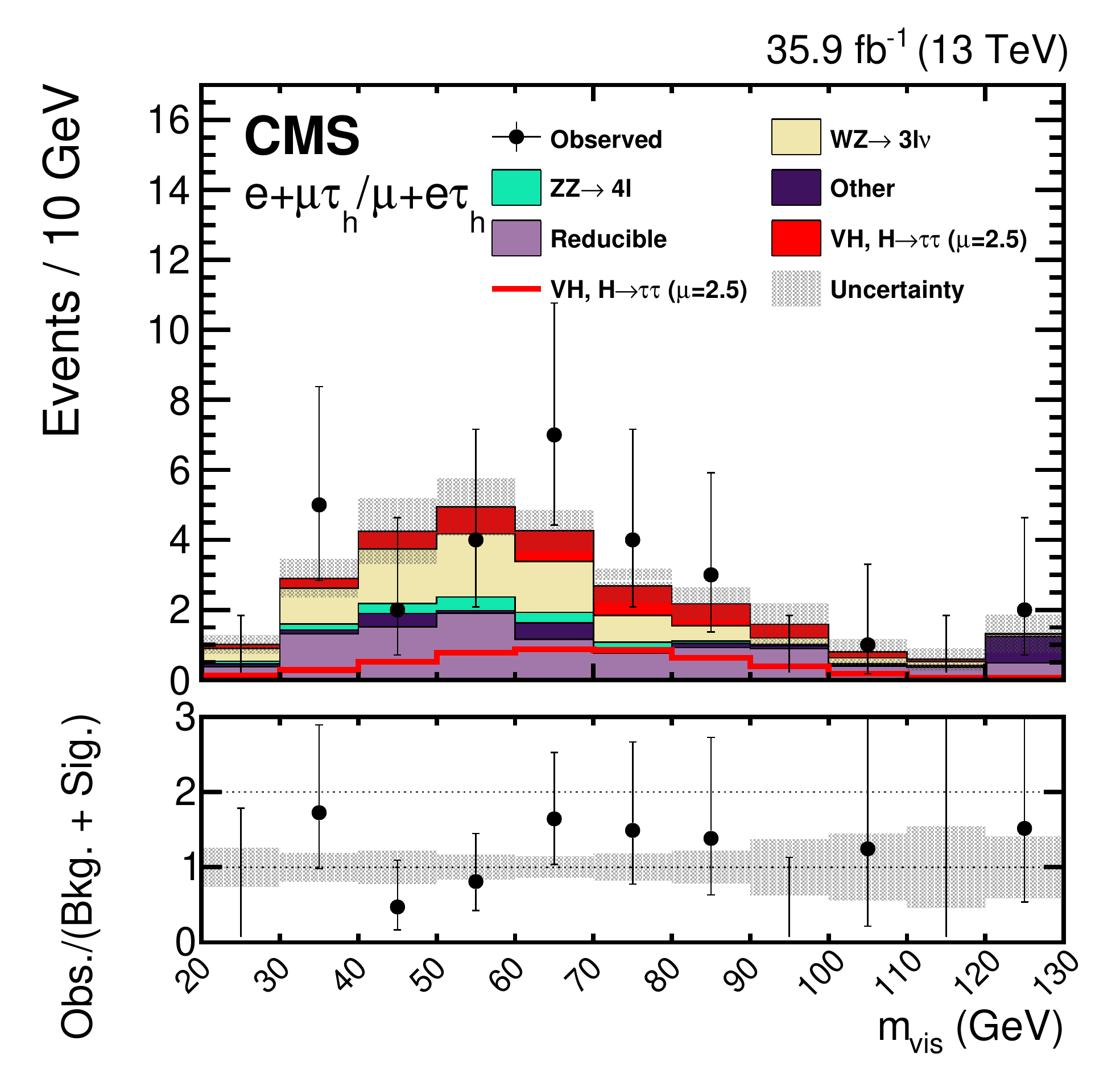}
  \includegraphics[width=0.45\textwidth]{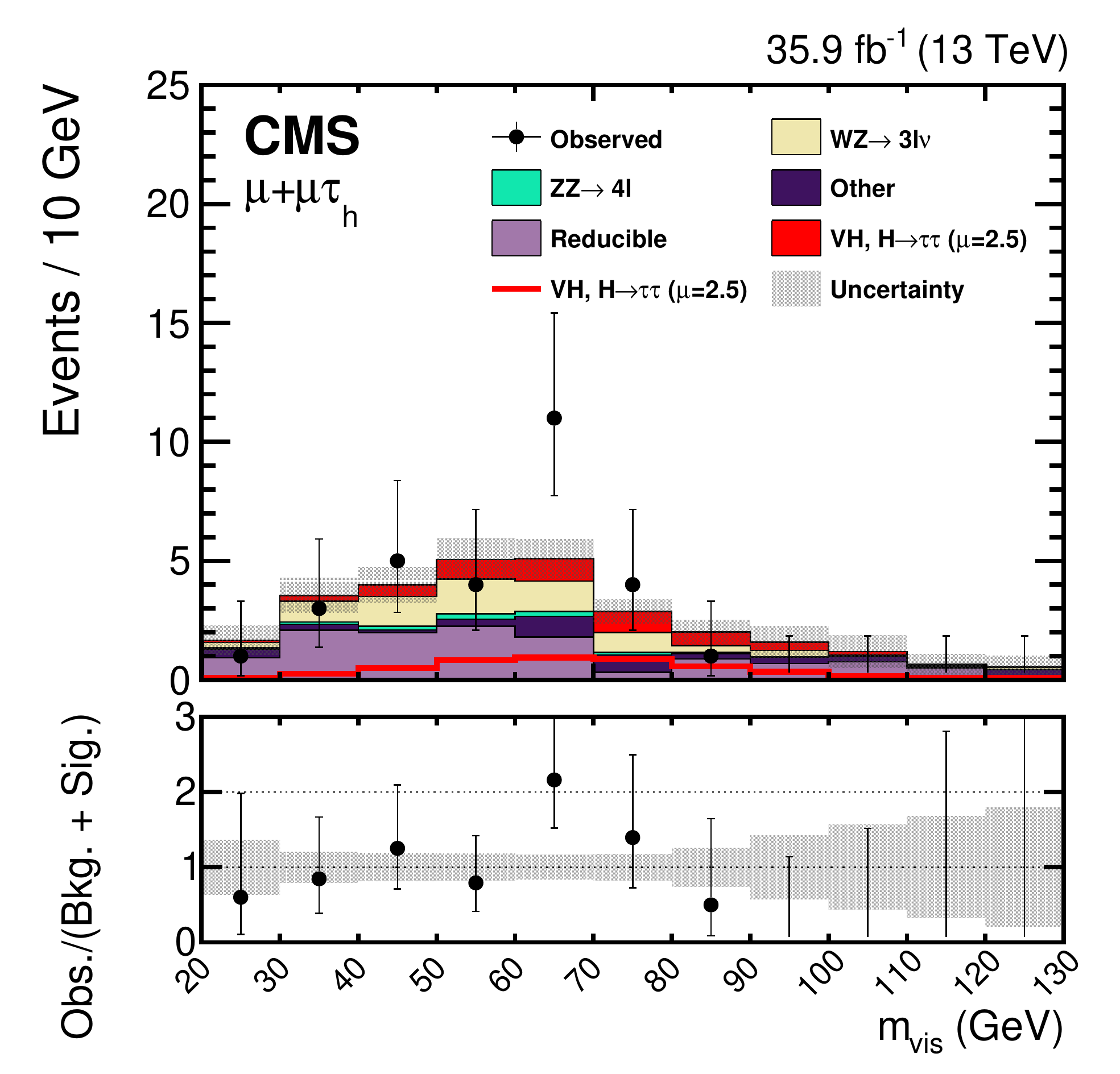} \\
  \includegraphics[width=0.45\textwidth]{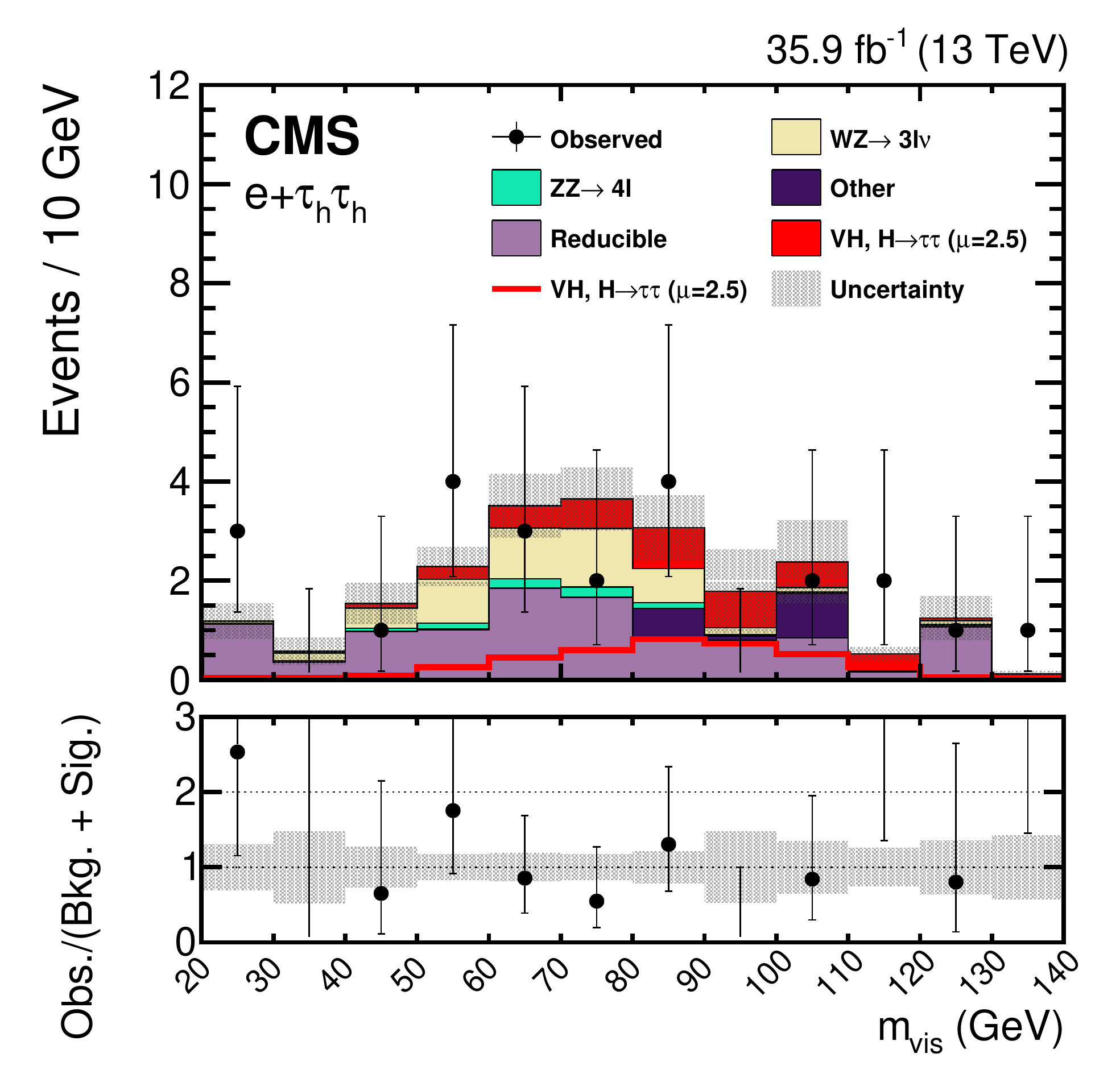}
  \includegraphics[width=0.45\textwidth]{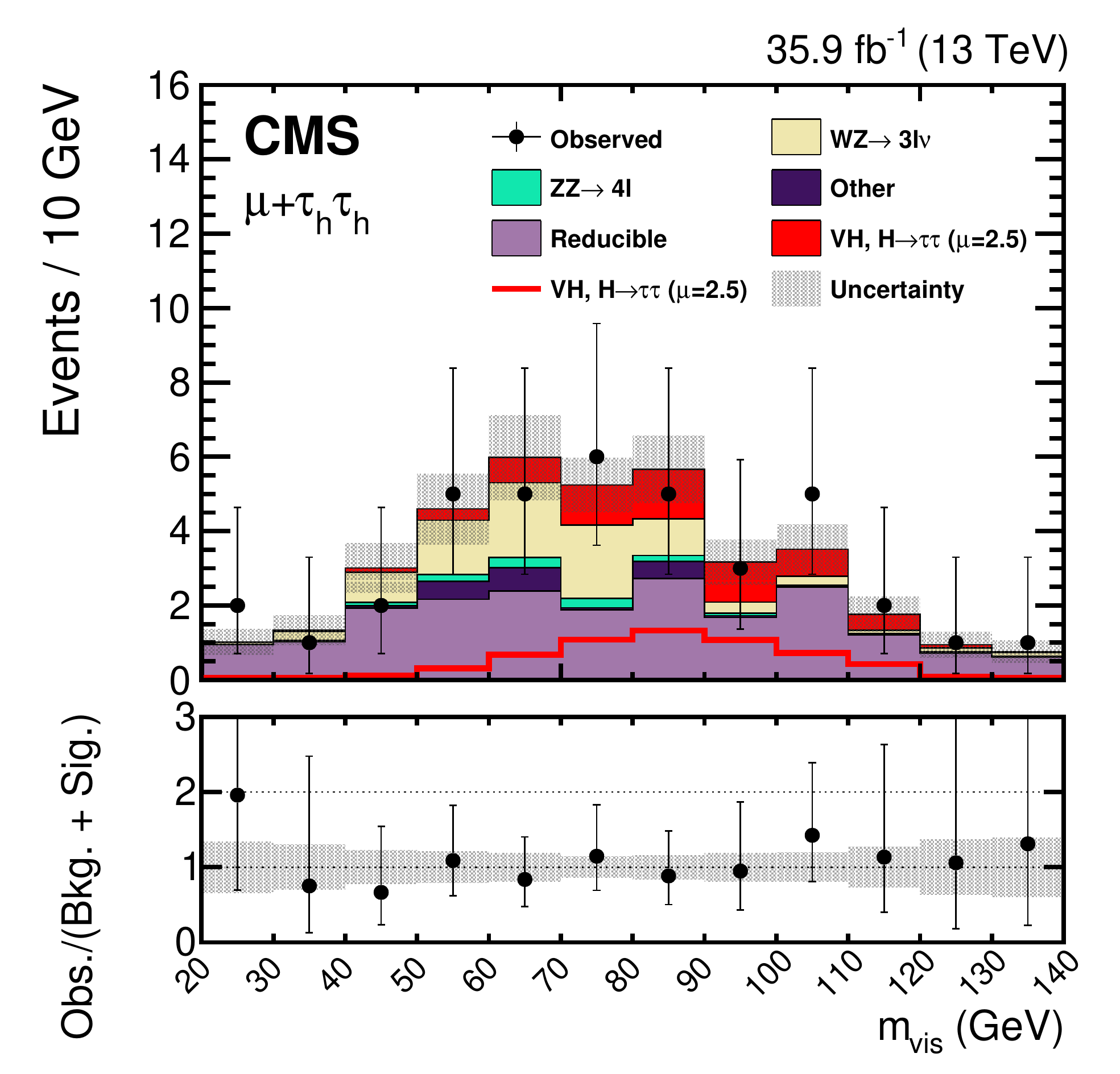}
 \caption{Post-fit visible mass distributions of the Higgs boson
  candidate in the $\Pe+\Pgm\tauh$/$\Pgm+\Pe\tauh$ (upper left),
  $\Pgm+\Pgm\tauh$ (upper right), $\Pe+\tauh\tauh$ (lower left),
  and $\Pgm+\tauh\tauh$ (lower right) final states.
  The uncertainties include both statistical and systematic components.
  The $\PW\PH$ and $\PZ\PH$, $\htt$ signal processes are summed
  together and shown as $\VH$, $\htt$ with a best fit $\mu = 2.5$.
  $\VH$, $\htt$ is shown both as a stacked filled histogram and an
  open overlaid histogram.
  The contribution from
  ``Other'' includes events from triboson, $\ttbar+\PW$/$\PZ$, $\ttbar\PH$ production,
  and all production modes leading to $\hww$ and $\hzz$ decays.
  In these distribution the $\PW\PH$, $\htt$ processes contributes
  91--93\% of the total of $\VH$, $\htt$.
 }
 \label{fig:mass_whs}
\end{figure}

\begin{figure}[h!t]
\centering
  \includegraphics[width=0.65\textwidth]{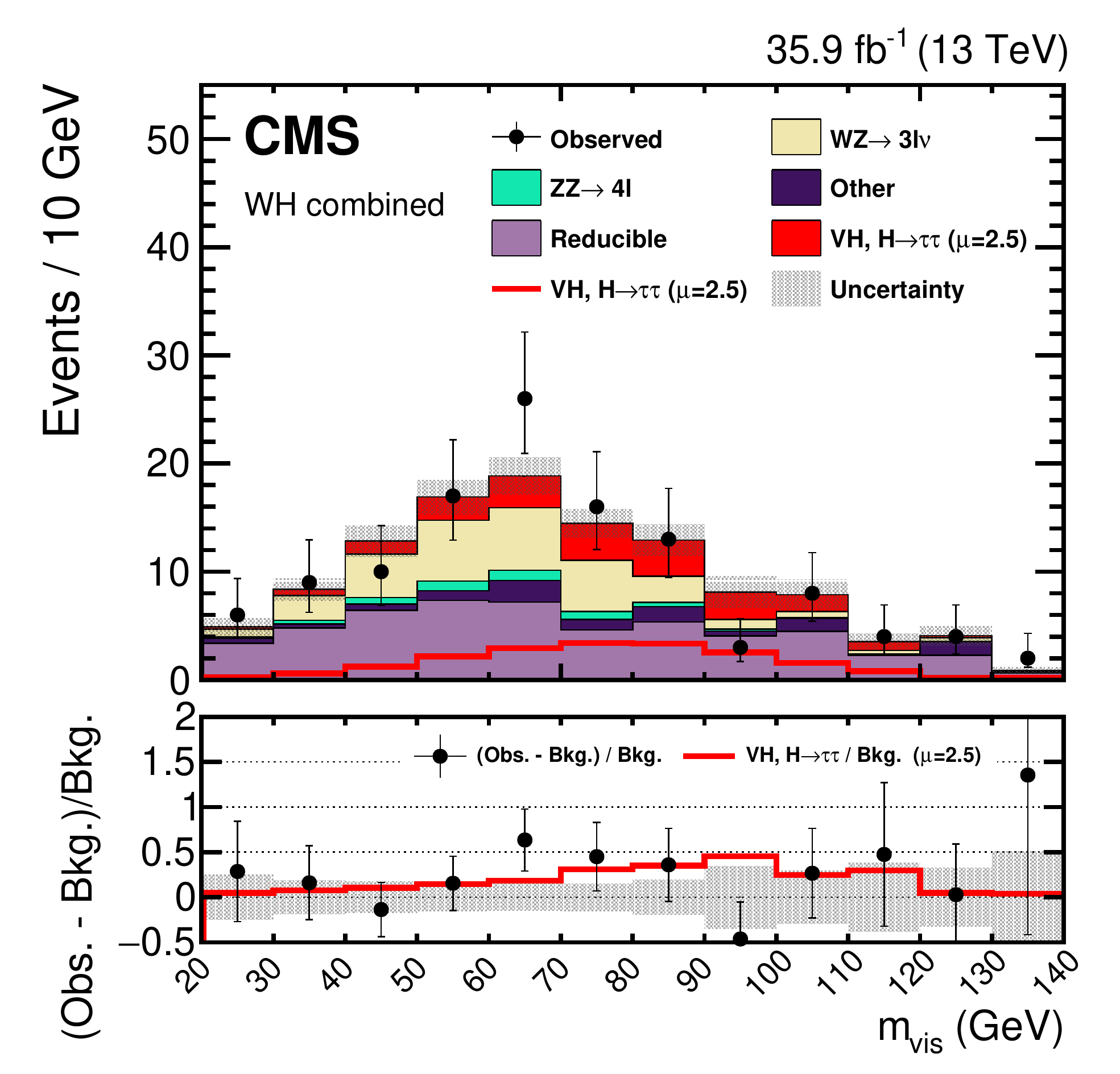}
 \caption{Post-fit visible mass distributions of the Higgs boson
  candidate in the four $\PW\PH$ final states
  combined together.
  The uncertainties include both statistical and systematic components.
  The $\PW\PH$ and $\PZ\PH$, $\htt$ signal processes are summed
  together and shown as $\VH$, $\htt$ with a best fit $\mu = 2.5$.
  $\VH$, $\htt$ is shown both as a stacked filled histogram and an
  open overlaid histogram.
  The contribution from
  ``Other'' includes events from triboson, $\ttbar+\PW$/$\PZ$, $\ttbar\PH$ production,
  and all production modes leading to $\hww$ and $\hzz$ decays.
  In this distribution the $\PW\PH$, $\htt$ process contributes
  92\% of the total of $\VH$, $\htt$.
 }
 \label{fig:mass_wh_mod}
\end{figure}

The results in the $\PW\PH$ channels are obtained from the distributions of the
visible mass of the $\tauh$ candidate pairs in the $\ell+\tauh\tauh$ channels,
and of the visible mass of the $\tauh$ and subleading light lepton in the
$\ell+\ell\tauh$ final states. The mass distributions
are shown in Fig.~\ref{fig:mass_whs} for the semileptonic
and hadronic channels. Figure~\ref{fig:mass_wh_mod} shows all
four $\PW\PH$ channels combined together.
The signal and background predicted yields, as well as the number of observed events,
are given for each final state for the $\PW\PH$ channel in Table~\ref{tab:sb_wh}.

\begin{table*}
\centering
\topcaption{Background and signal expectations for the $\PZ\PH$ channels,
together with the numbers of observed
events, for the post-fit signal region distributions. The $\PZ\PH$ final states
are each grouped according to the Higgs boson decay products.
The $\ell\ell$ notation covers both $\PZ \to \Pgm\Pgm$ and $\PZ \to \Pe\Pe$ events.
The $\PW\PH$ and $\PZ\PH$, $\htt$ signal yields are listed both individually and summed
together, and correspond to $\htt$ with a best fit $\mu = 2.5$ for a Higgs boson with a mass $\mH = 125\GeV$.
The background uncertainty accounts for all sources of background uncertainty,
systematic as well as statistical, after the global fit. The contribution from
``Other'' includes events from triboson, $\ttbar+\PW$/$\PZ$, $\ttbar\PH$ production,
and all production modes leading to $\hww$ and $\hzz$ decays.
}
\label{tab:sb_zh}
\newcolumntype{x}{D{,}{\,\pm\,}{5.5}}
\begin{tabular}{lxxxx}
Process & \multicolumn{1}{c}{$\ell\ell+\Pe\tauh$} &  \multicolumn{1}{c}{$\ell\ell+\Pgm\tauh$} &  \multicolumn{1}{c}{$\ell\ell+\tauh\tauh$} &  \multicolumn{1}{c}{$\ell\ell+\Pe\Pgm$} \\
\hline
$\PZ\PZ$                        & 14.40, 0.36 & 26.91, 0.55 & 25.58, 1.05 & 9.33, 0.18 \\
Reducible                       & 14.01, 1.55 & 17.58, 1.17 & 58.05, 2.87 & 3.66, 4.60 \\
Other                           & 0.62, 0.08  & 1.54, 0.61  & 0.81, 0.42  & 3.02, 0.23 \\
Total backgrounds               & 29.03, 1.59 & 46.03, 1.43 & 84.44, 3.08 & 16.01, 4.61\\[\cmsTabSkip]
$\PW\PH, \PH \to\PGt\PGt$       & 0.008, 0.002  & 0.010, 0.003  & 0.016, 0.005  & 0.002, 0.001 \\
$\PZ\PH, \PH \to\PGt\PGt$       & 2.83, 0.39  & 5.31, 0.70  & 5.29, 1.17  & 1.62, 0.20 \\
Total signal                    & 2.84, 0.39  & 5.32, 0.70  & 5.31, 1.17  & 1.62, 0.20 \\[\cmsTabSkip]
Observed &  \multicolumn{1}{c}{33} &  \multicolumn{1}{c}{53} &  \multicolumn{1}{c}{87} &  \multicolumn{1}{c}{20}  \\
\end{tabular}
\end{table*}

\begin{table*}
\centering
\topcaption{Background and signal expectations for the $\PW\PH$ channels,
together with the numbers of observed
events, for the post-fit signal region distributions.
The $\PW\PH$ and $\PZ\PH$, $\htt$ signal yields are listed both individually and summed
together, and correspond to $\htt$ with a best fit $\mu = 2.5$ for a Higgs boson with a mass $\mH = 125\GeV$.
The background uncertainty accounts for all sources of background uncertainty,
systematic as well as statistical, after the global fit. The contributions from
triboson, $\ttbar+\PW$/$\PZ$, $\ttbar\PH$ production,
and all production modes leading to $\hww$ and $\hzz$ decays
are included in the category labeled ``Other''.
}
\label{tab:sb_wh}
\newcolumntype{x}{D{,}{\,\pm\,}{5.5}}
\begin{tabular}{lxxxx}
Process & \multicolumn{1}{c}{$\Pe+\Pgm\tauh$/$\Pgm+\Pe\tauh$ } & \multicolumn{1}{c}{$\Pgm+\Pgm\tauh$ } & \multicolumn{1}{c}{$\Pe+\tauh\tauh$} & \multicolumn{1}{c}{$\Pgm+\tauh\tauh$}  \\
\hline
$\PZ\PZ$                  & 1.56, 0.05    & 0.93, 0.03  & 0.82, 0.04  & 1.18, 0.05   \\
$\PW\PZ$                  & 7.92, 0.28    & 6.69, 0.24  & 4.83, 0.25  & 8.38, 0.42   \\
Reducible                 & 10.09, 1.61   & 12.19, 1.72 & 10.68, 1.27 & 19.80, 1.87  \\
Other                     & 2.28, 0.61    & 3.77, 0.84  & 1.71, 1.08  & 1.76, 0.90   \\
Total backgrounds         & 21.85, 1.75   & 23.58, 1.93 & 18.04, 1.69 & 31.12, 2.12  \\[\cmsTabSkip]
$\PW\PH, \PH \to\PGt\PGt$ & 4.28, 0.72    & 4.25, 0.73  & 3.51, 0.62  &  5.45, 0.97  \\
$\PZ\PH, \PH \to\PGt\PGt$ & 0.42, 0.07    & 0.40, 0.08  & 0.33, 0.07  &  0.44, 0.10  \\
Total signal              & 4.70, 0.72    & 4.65, 0.73  & 3.84, 0.62  &  5.89, 0.98  \\[\cmsTabSkip]
Observed &  \multicolumn{1}{c}{28} &  \multicolumn{1}{c}{29} &  \multicolumn{1}{c}{23} &  \multicolumn{1}{c}{38}  \\
\end{tabular}
\end{table*}

\begin{figure}[!ht]
 \centering
  \includegraphics[width=0.65\textwidth]{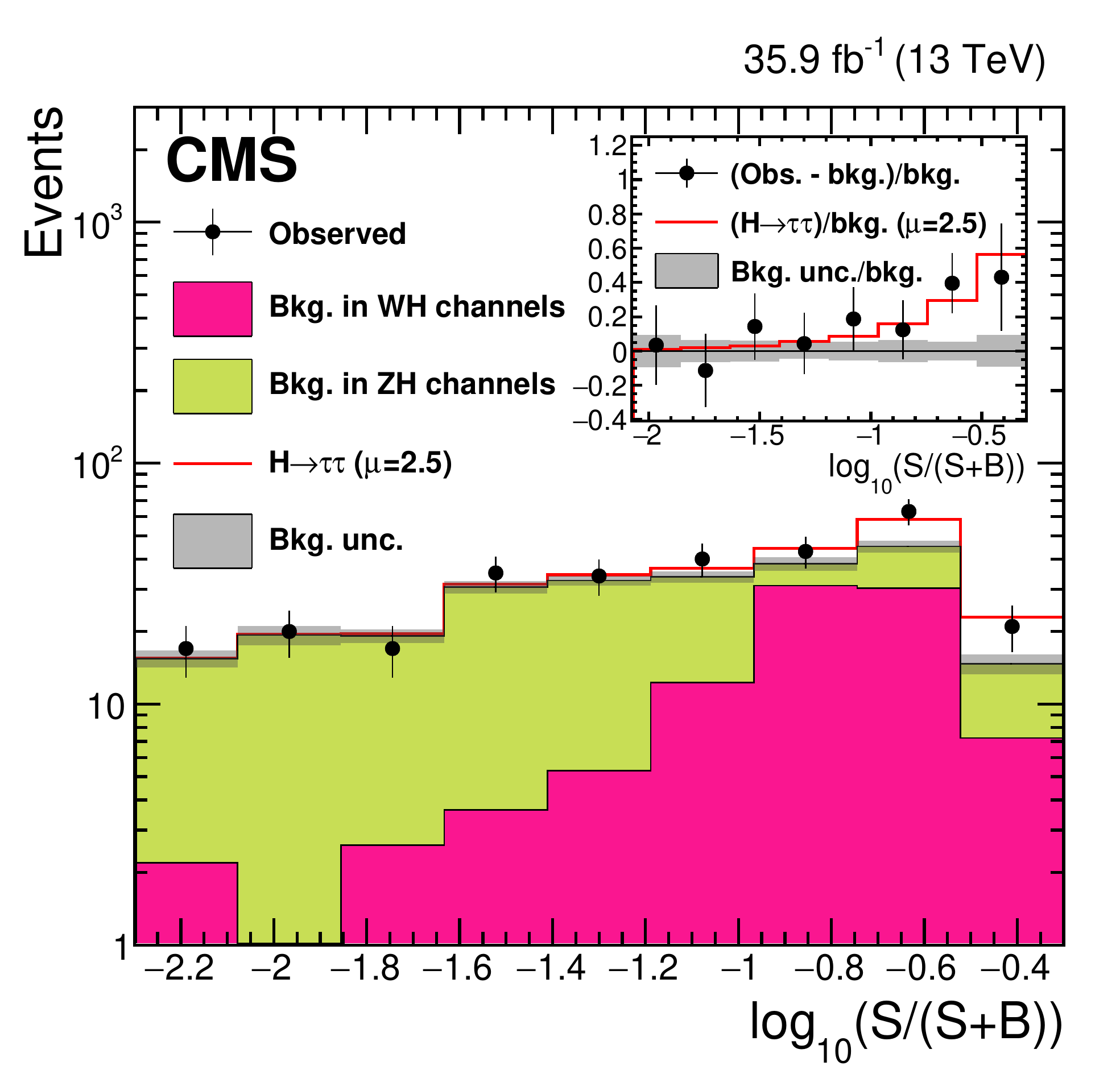}
 \caption{
 Distribution of the decimal logarithm of the ratio between the expected signal and the
 sum of the expected signal and background. The signal, corresponding
 to the best fit value $\mu=2.5$, and
 expected background in each bin of the mass distributions
 used to extract the results, in all final states are combined.
 The background contributions are
 separated based on the analysis channel, $\PW\PH$ or $\PZ\PH$. The inset
 shows the corresponding difference between the
 data and expected background distributions divided by the background expectation,
 as well as the signal expectation divided by the background expectation.
 }
 \label{fig:sb}
\end{figure}

Events from all final states are combined as a function of their decimal logarithm of the ratio of the
signal ($S$) to signal-plus-background ($S+B$) in each bin, as shown in Fig.~\ref{fig:sb}.
Most of the $\PZ\PH$ and $\PW\PH$ final states contribute to the most sensitive bins in this distribution.
The sensitive bins in the mass distributions correspond to those that include the peak of the signal
from approximately 70--110\GeV in the $\mvis$ distributions from the $\PW\PH$ channels and
100--160\GeV in the $\mtt$ distributions from the $\PZ\PH$ channels. The least sensitive bins in
Fig.~\ref{fig:sb} include background events from all channels away from the signal peak and especially in the low
$\LTH$ region for the $\PZ\PH$ channels.
An excess of observed events with respect to the SM background expectation is
visible in the most sensitive bins of the analysis.

The maximum likelihood fit to the $\PW\PH$ and $\PZ\PH$ associated production event distributions yields a signal strength
$\mu = 2.5 ^{+1.4} _{-1.3}$ ($1.0 ^{+1.1} _{-1.0}$ expected)
for a significance of 2.3 standard deviations (1.0 expected).
The large $\mu$ value is
driven by the $\PW\PH$ channels, where the observation significantly exceeds the expectations from the SM including the Higgs boson.
The constraints from the combined global fit are used to extract the
individual best fit signal strengths for $\PW\PH$ and $\PZ\PH$:
$\mu_{\PW\PH} = 3.6^{+1.8} _{-1.6}$ ($1.0 ^{+1.6} _{-1.4}$ expected), and
$\mu_{\PZ\PH} = 1.4^{+1.6} _{-1.5}$ ($1.0 ^{+1.5} _{-1.3}$ expected).

The results
of this dedicated $\PW\PH$ and $\PZ\PH$ associated production analysis are combined with the prior
$\PH \to \PGt\PGt$ analysis that targeted the gluon fusion and
vector boson fusion production modes using the same data set and dilepton final states~\cite{Sirunyan:2017khh}.
The signal regions in both analyses are orthogonal by design because events with extra leptons
are removed from the gluon fusion and vector boson fusion targeted dilepton final states.
Changes in the gluon fusion signal modeling and uncertainties were made between
the publication of Ref.~\cite{Sirunyan:2017khh} and the combination presented here, to take advantage of the most accurate, available
simulations of the gluon fusion process.
The gluon fusion simulation used in Ref.~\cite{Sirunyan:2017khh} was computed with next-to-leading order matrix
elements merged with the parton shower ($\textrm{NLO}+\textrm{PS}$) accuracy.
These $\textrm{NLO}+\textrm{PS}$ gluon fusion samples were reweighted to match the Higgs boson
$\pt$ spectrum from the \textsc{nnlops} generator~\cite{Hamilton:2013fea}. Additionally, the
gluon fusion cross section uncertainty scheme has been updated to the one proposed in Ref.~\cite{deFlorian:2016spz}.
This uncertainty scheme includes 9 nuisance parameters accounting for the uncertainties in the cross
section prediction for exclusive jet bins, the 2-jet and 3-jet VBF phase space regions, different Higgs boson $\pt$
regions, and the uncertainty in the Higgs boson $\pt$ distribution due to missing higher-order corrections
relating to the treatment of the top quark mass.

After applying the mentioned changes to the gluon fusion modeling, the gluon fusion and VBF targeted
analysis results in a best fit signal strength for $\htt$ of $\mu = 1.17 ^{+0.27} _{-0.25}$
($1.00 ^{+0.25} _{-0.23}$ expected).

\begin{figure}[!ht]
 \centering
  \includegraphics[width=0.65\textwidth]{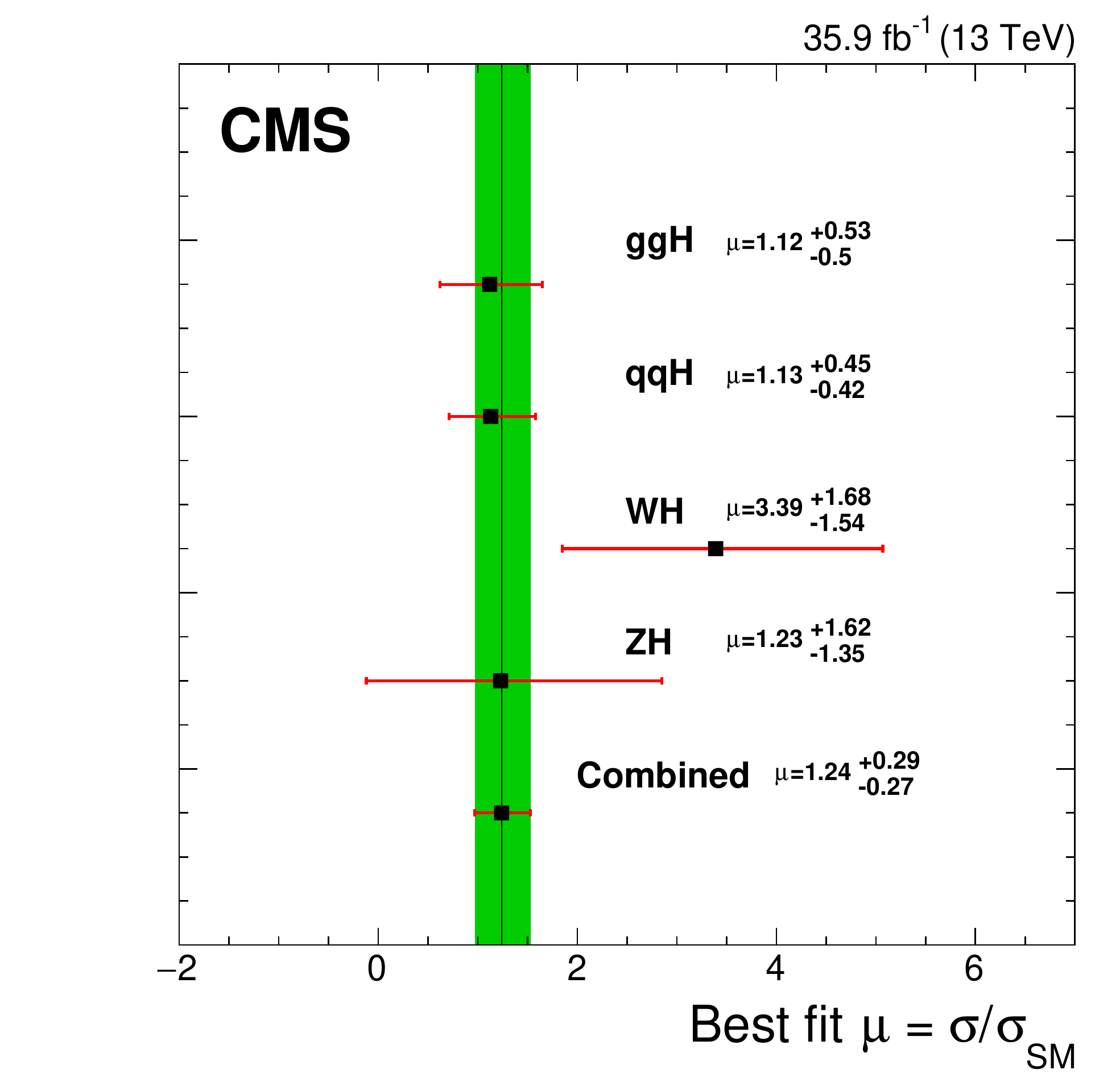}
 \caption{
 Best fit signal strength per Higgs boson production process, for $\mH = 125$\GeV,
 using a combination of the $\PW\PH$ and $\PZ\PH$ targeted analysis detailed in this paper
 with the CMS analysis performed in the same data set for the same decay mode but targeting
 the gluon fusion and vector boson fusion production mechanisms~\cite{Sirunyan:2017khh}.
 The constraints from the combined global fit are used to extract each of the
 individual best fit signal strengths. The combined best fit signal strength
 is $\mu = 1.24 ^{+0.29} _{-0.27}$.
 }
 \label{fig:mu_higgs_processes}
\end{figure}

With combined results, the significance, signal strengths, and Higgs boson couplings
can be measured with better precision than with either analysis alone.
The combination leads to an
observed significance of 5.5 standard deviations (4.8 expected).
The best fit signal strength for the combination is $\mu = 1.24 ^{+0.29} _{-0.27}$
($1.00 ^{+0.24} _{-0.23}$ expected).
The signal regions used in the combination target the four leading Higgs boson production
mechanisms allowing extraction of the Higgs boson signal strength per
production mechanism.
The production mode specific signal strength measurements are shown in Fig.~\ref{fig:mu_higgs_processes}.

\begin{figure}[!ht]
 \centering
  \includegraphics[width=0.65\textwidth]{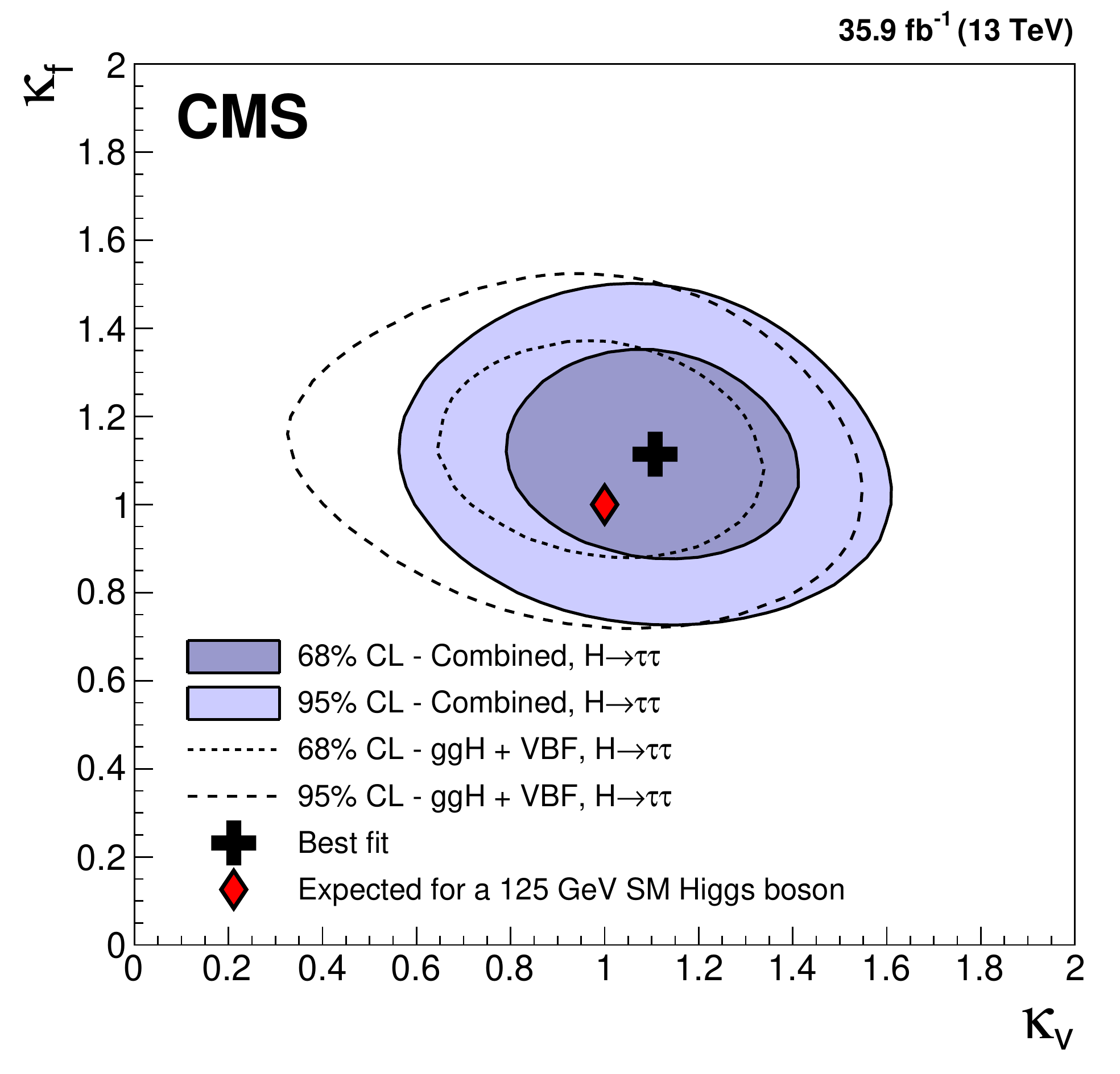}
 \caption{Scans of the negative
 log-likelihood difference as a function of $\kappa_V$ and $\kappa_f$, for
 $\mH = 125\GeV$. Contours corresponding to confidence levels (\CL) of 68 and 95\% are shown.
 All nuisance parameters are profiled for each point.
 The scan labeled as ``Combined'' is a combination of the $\PW\PH$ and $\PZ\PH$ targeted analysis detailed in this paper
 with the CMS analysis performed in the same data set for the same decay mode but targeting the
 gluon fusion and vector boson fusion production mechanisms~\cite{Sirunyan:2017khh}.
 The results for the gluon fusion and vector boson fusion analysis are represented by the dashed lines
 and are labeled as ``$\Pg\Pg\PH+\textrm{VBF}$''.
 For these scans, the included $\hww$ and $\hzz$ processes
 are treated as signal.
 }
 \label{fig:kVkf}
\end{figure}

This combination places a tighter constraint on the
$\PH \to \PGt\PGt$ process in the $(\kappa_\text{V}$,$\kappa_\text{f})$ Higgs boson couplings parameter space
than previous analyses targeting exclusively the $\htt$ decay process.
The coupling parameters $\kappa_\text{V}$ and $\kappa_\text{f}$ quantify, respectively,
the ratio between the measured and the SM expected values for the couplings of the Higgs boson to
vector bosons and to fermions, with the methods described in Ref.~\cite{Chatrchyan:2014nva}.
Constraints are set with a likelihood scan that is performed for $\mH=125\GeV$ in the ($\kappa_\text{V}$,$\kappa_\text{f}$)
parameter space.
For this scan only, Higgs boson decays to pairs of $\PW$ or $\PZ$ bosons, $\hww$ or $\hzz$,
are considered as part of the signal. All nuisance
parameters are profiled for each point of the scan. As shown in
Fig.~\ref{fig:kVkf}, the observed likelihood contour is consistent with the SM expectations
of $\kappa_\text{V}$ and $\kappa_\text{f}$ equal to unity providing increased confidence that the
Higgs boson couples to $\PGt$ leptons through a Yukawa coupling as predicted in the SM.
The addition of the $\PW\PH$ and $\PZ\PH$ targeted final states brings roughly a 10\% reduction
in the maximum extent of the 68\% \CL for $\kappa_\text{V}$ compared to the gluon fusion and
vector boson fusion targeted analysis.

\section{Summary}

A search is presented for the standard model (SM) Higgs boson in $\PW\PH$ and $\PZ\PH$ associated production
processes, based on data collected in proton-proton collisions by the
CMS detector in 2016 at a center-of-mass energy of 13\TeV.
Event categories are defined by three-lepton final states targeting $\PW\PH$ production,
and four-lepton final states targeting $\PZ\PH$ production.
The best fit signal
strength is $\mu = 2.5 ^{+1.4} _{-1.3}$ ($1.0 ^{+1.1} _{-1.0}$ expected)
for a significance of 2.3 standard deviations (1.0 expected).

The results of this analysis are combined with those of the CMS analysis targeting
gluon fusion and vector boson fusion production, also performed at a center-of-mass energy of 13\TeV,
and constraints on the $\htt$ decay rate are set.
The best fit signal strength is $\mu = 1.24 ^{+0.29} _{-0.27}$
($1.00 ^{+0.24} _{-0.23}$ expected), and the
observed significance is 5.5 standard deviations (4.8 expected) for a Higgs boson mass of 125\GeV.
This combination further constrains the coupling
of the Higgs boson to vector bosons, resulting in measured couplings
that are consistent with SM predictions within one standard deviation, providing increased
confidence that the Higgs boson couples to $\PGt$ leptons through a Yukawa coupling as predicted in the SM.
The combination allows for extraction of the signal strengths for the four leading Higgs boson
production processes using exclusively $\htt$ targeted final states, the results of which are
largely consistent with the SM.
The measurements of the Higgs boson production mechanisms using $\htt$ decays
are the best results to date for the $\PW\PH$ and $\PZ\PH$ associated production mechanisms using the $\htt$ process.

\begin{acknowledgments}
We congratulate our colleagues in the CERN accelerator departments for the excellent performance of the LHC and thank the technical and administrative staffs at CERN and at other CMS institutes for their contributions to the success of the CMS effort. In addition, we gratefully acknowledge the computing centers and personnel of the Worldwide LHC Computing Grid for delivering so effectively the computing infrastructure essential to our analyses. Finally, we acknowledge the enduring support for the construction and operation of the LHC and the CMS detector provided by the following funding agencies: BMBWF and FWF (Austria); FNRS and FWO (Belgium); CNPq, CAPES, FAPERJ, FAPERGS, and FAPESP (Brazil); MES (Bulgaria); CERN; CAS, MoST, and NSFC (China); COLCIENCIAS (Colombia); MSES and CSF (Croatia); RPF (Cyprus); SENESCYT (Ecuador); MoER, ERC IUT, and ERDF (Estonia); Academy of Finland, MEC, and HIP (Finland); CEA and CNRS/IN2P3 (France); BMBF, DFG, and HGF (Germany); GSRT (Greece); NKFIA (Hungary); DAE and DST (India); IPM (Iran); SFI (Ireland); INFN (Italy); MSIP and NRF (Republic of Korea); MES (Latvia); LAS (Lithuania); MOE and UM (Malaysia); BUAP, CINVESTAV, CONACYT, LNS, SEP, and UASLP-FAI (Mexico); MOS (Montenegro); MBIE (New Zealand); PAEC (Pakistan); MSHE and NSC (Poland); FCT (Portugal); JINR (Dubna); MON, RosAtom, RAS, RFBR, and NRC KI (Russia); MESTD (Serbia); SEIDI, CPAN, PCTI, and FEDER (Spain); MOSTR (Sri Lanka); Swiss Funding Agencies (Switzerland); MST (Taipei); ThEPCenter, IPST, STAR, and NSTDA (Thailand); TUBITAK and TAEK (Turkey); NASU and SFFR (Ukraine); STFC (United Kingdom); DOE and NSF (USA).

\hyphenation{Rachada-pisek} Individuals have received support from the Marie-Curie program and the European Research Council and Horizon 2020 Grant, contract No. 675440 (European Union); the Leventis Foundation; the A. P. Sloan Foundation; the Alexander von Humboldt Foundation; the Belgian Federal Science Policy Office; the Fonds pour la Formation \`a la Recherche dans l'Industrie et dans l'Agriculture (FRIA-Belgium); the Agentschap voor Innovatie door Wetenschap en Technologie (IWT-Belgium); the F.R.S.-FNRS and FWO (Belgium) under the ``Excellence of Science - EOS" - be.h project n. 30820817; the Ministry of Education, Youth and Sports (MEYS) of the Czech Republic; the Lend\"ulet (``Momentum") Programme and the J\'anos Bolyai Research Scholarship of the Hungarian Academy of Sciences, the New National Excellence Program \'UNKP, the NKFIA research grants 123842, 123959, 124845, 124850 and 125105 (Hungary); the Council of Science and Industrial Research, India; the HOMING PLUS program of the Foundation for Polish Science, cofinanced from European Union, Regional Development Fund, the Mobility Plus program of the Ministry of Science and Higher Education, the National Science Center (Poland), contracts Harmonia 2014/14/M/ST2/00428, Opus 2014/13/B/ST2/02543, 2014/15/B/ST2/03998, and 2015/19/B/ST2/02861, Sonata-bis 2012/07/E/ST2/01406; the National Priorities Research Program by Qatar National Research Fund; the Programa Estatal de Fomento de la Investigaci{\'o}n Cient{\'i}fica y T{\'e}cnica de Excelencia Mar\'{\i}a de Maeztu, grant MDM-2015-0509 and the Programa Severo Ochoa del Principado de Asturias; the Thalis and Aristeia programs cofinanced by EU-ESF and the Greek NSRF; the Rachadapisek Sompot Fund for Postdoctoral Fellowship, Chulalongkorn University and the Chulalongkorn Academic into Its 2nd Century Project Advancement Project (Thailand); the Welch Foundation, contract C-1845; and the Weston Havens Foundation (USA).
\end{acknowledgments}

\clearpage

\bibliography{auto_generated}

\cleardoublepage \appendix\section{The CMS Collaboration \label{app:collab}}\begin{sloppypar}\hyphenpenalty=5000\widowpenalty=500\clubpenalty=5000\vskip\cmsinstskip
\textbf{Yerevan Physics Institute, Yerevan, Armenia}\\*[0pt]
A.M.~Sirunyan, A.~Tumasyan
\vskip\cmsinstskip
\textbf{Institut f\"{u}r Hochenergiephysik, Wien, Austria}\\*[0pt]
W.~Adam, F.~Ambrogi, E.~Asilar, T.~Bergauer, J.~Brandstetter, M.~Dragicevic, J.~Er\"{o}, A.~Escalante~Del~Valle, M.~Flechl, R.~Fr\"{u}hwirth\cmsAuthorMark{1}, V.M.~Ghete, J.~Hrubec, M.~Jeitler\cmsAuthorMark{1}, N.~Krammer, I.~Kr\"{a}tschmer, D.~Liko, T.~Madlener, I.~Mikulec, N.~Rad, H.~Rohringer, J.~Schieck\cmsAuthorMark{1}, R.~Sch\"{o}fbeck, M.~Spanring, D.~Spitzbart, A.~Taurok, W.~Waltenberger, J.~Wittmann, C.-E.~Wulz\cmsAuthorMark{1}, M.~Zarucki
% ...
\vskip\cmsinstskip
% ...
1:  Also at Vienna University of Technology, Vienna, Austria\\
% ...
\end{sloppypar}
\end{document}